\documentclass[10pt,conference]{IEEEtran}
\usepackage{mathtools}

\usepackage{lipsum}
\usepackage{tikz}     
\usepackage{pifont} 

\usepackage{tabularx} 

\usepackage{booktabs}
\usepackage{siunitx}

\usepackage{amsmath}
\usepackage{amsfonts,amssymb,amsbsy}

\IEEEoverridecommandlockouts        
\usepackage{graphics}
\usepackage{epsfig} 
\usepackage{siunitx} 
\usepackage{array}   
\usepackage[super]{natbib}
\usepackage{comment}

\usepackage{xcolor} 
\usepackage{caption}

\usepackage{mathptmx} 
\usepackage{times} 
\usepackage{array}
\usepackage{hyperref}
\usepackage{stmaryrd}  

\usepackage{blindtext, graphicx}   
\usepackage[tc]{titlepic}   
\usepackage{float}   
\usepackage{cuted}     
\usepackage{amsthm}   
\newtheoremstyle{case}{}{}{}{}{}{:}{ }{}
\usepackage{authblk}
\usepackage{caption}
\usepackage{subcaption}

\usepackage{xcolor}
\usepackage{soul}

\begin{document}

\title{RAPID-Net: Accurate Pocket Identification for Binding-Site-Agnostic Docking}

\date{\vspace{-5ex}}

\author[*,1]{Yaroslav Balytskyi}
\author[2]{Inna Hubenko}
\author[2]{Alina Balytska}
\author[1,3]{Christopher V. Kelly}

\affil[1]{Department of Physics and Astronomy, Wayne State University, Detroit, MI, 48201, USA}

\affil[2]{Cherkasy Medical Academy, Cherkasy, 18000, Ukraine}

\affil[3]{Center for Integrative Metabolic and Endocrine Research, School of Medicine, Wayne State University, Detroit, MI, 48201, USA}

\maketitle

\begin{abstract}

Accurate identification of druggable pockets and their features is essential for structure-based drug design and effective downstream docking. Here, we present RAPID-Net, a deep learning-based algorithm designed for the accurate prediction of binding pockets and seamless integration with docking pipelines. On the PoseBusters benchmark, RAPID-Net–guided AutoDock Vina achieves 54.9\% of Top-1 poses with RMSD~$<2\,\text{\AA}$ and satisfying the PoseBusters chemical‐validity criterion, compared to 49.1\% for DiffBindFR. On the most challenging time split of PoseBusters aiming to assess generalization ability (structures submitted after September 30, 2021), RAPID-Net-guided AutoDock Vina achieves 53.1\% of Top-1 poses with RMSD~$<2\,\text{\AA}$ and PB-valid, versus 59.5\% for AlphaFold 3. Notably, in 92.2\% of cases, RAPID-Net-guided Vina samples at least one pose with RMSD~$<2\,\text{\AA}$ (regardless of its rank), indicating that pose ranking, rather than sampling, is the primary accuracy bottleneck. The lightweight inference, scalability, and competitive accuracy of RAPID-Net position it as a viable option for large-scale virtual screening campaigns. Across diverse benchmark datasets, RAPID-Net outperforms other pocket prediction tools, including PUResNet and Kalasanty, in both docking accuracy and pocket–ligand intersection rates. Furthermore, we demonstrate the potential of RAPID-Net to accelerate the development of novel therapeutics by highlighting its performance on pharmacologically relevant targets. RAPID-Net accurately identifies distal functional sites, offering new opportunities for allosteric inhibitor design. In the case of the RNA-dependent RNA polymerase of SARS-CoV-2, RAPID-Net uncovers a wider array of potential binding pockets than existing predictors, which typically annotate only the orthosteric pocket and overlook secondary cavities.

RAPID-Net, along with interactive demonstration notebooks, is publicly available at: \href{https://github.com/BalytskyiJaroslaw/RAPID-Net}{\texttt{github.com/BalytskyiJaroslaw/RAPID-Net}}, offering a user-friendly and fully reproducible framework for pocket-guided docking.

*Corresponding Author: hr6998@wayne.edu
\end{abstract}


\section{Introduction and Problem Statement}

Molecular docking is the computational problem of predicting the most favorable ligand binding poses in a protein-ligand complex, given the experimentally determined or computer-simulated protein structure and the initial conformation of the ligand~\cite{kukol2008molecular}. It plays a key role in drug development, typically following target identification when the biological molecule responsible for a disease is identified and validated for therapeutic intervention~\cite{an2004comprehensive}. Once the target is identified, molecular docking is used for structure-based drug discovery and development~\cite{de2016molecular,rezaei2020deep} to identify and optimize potential lead compounds~\cite{vamathevan2019applications,saikia2019molecular}. The docking results can guide virtual screening workflows, enabling the selection of promising candidates from vast chemical libraries~\cite{zoete2009docking,lionta2014structure}. This process helps to determine whether a compound has potential for drug development~\cite{patrick2023introduction}. By significantly improving the speed and efficiency of early drug discovery, molecular docking reduces the time and costs associated with the development of new therapeutics~\cite{hernandez2013protein}.

Typical protein-ligand docking pipelines rely on users specifying the binding pocket, and docking programs such as AutoDock Vina~\cite{trott2010autodock,eberhardt2021autodock}, GOLD~\cite{verdonk2003improved}, and Glide~\cite{friesner2004glide} use grids to confine their search to known or hypothesized protein's interaction sites. However, in the absence of such information -- in ``blind'' or binding-site-agnostic settings -- protein-ligand docking becomes significantly more difficult~\cite{grasso2022fragmented}, as the docking algorithm must scan the entire protein surface, dramatically increasing the computational complexity. Traditional blind docking methods often rely on extensive sampling to explore potential binding sites across the whole protein, but this approach is computationally and time-intensive, making it impractical for large-scale virtual screening tasks~\cite{hassan2017protein,koes2013lessons}. With the advent of advanced protein structure prediction methods~\cite{shen2022predicting}--including AlphaFold~\cite{jumper2021highly}, ColabFold~\cite{mirdita2022colabfold}, OpenFold~\cite{ahdritz2024openfold}, and RosettaFold~\cite{baek2021accurate}--the number of protein structures generated has surged, often without any corresponding ligand information. Meanwhile, only about $\sim 5\%$ of the human coding genes currently serve as commercial drug targets, leaving a wide range of disease-related targets unexplored~\cite{overington2006many}. Despite this immense potential, most new drug research and development remain focused on a limited set of well-established targets, underscoring the urgent need for novel druggable targets~\cite{santos2017comprehensive}. As more proteins lacking known binding pocket information are considered in drug discovery~\cite{huang2023dsdp}, the demand for ligand docking approaches that do not rely on prior knowledge of the binding site has grown significantly, highlighting the necessity for binding-site-agnostic methods.

In this work, we aim to develop a high‐accuracy pocket-finding algorithm (\textbf{RAPID-Net}) optimized for seamless integration with standard docking pipelines to deliver efficient guidance in binding-site-agnostic workflows. Many protein binding site predictors are benchmarked solely on the geometric properties of their predicted pockets~\cite{utges2024comparative}, but their impact on the accuracy of downstream ligand docking is rarely assessed. While many factors affect docking performance, we show further in the text that pocket prediction quality has a decisive role in determining docking success. We evaluate RAPID-Net's performance both by docking accuracy -- using the widely adopted AutoDock Vina~\cite{eberhardt2021autodock} -- and by pocket-ligand intersection rates, as detailed further in the text. Indeed, RAPID-Net can provide the search grid centers and dimensions for any docking engine, and Vina was chosen due to its popularity.

Despite relying on Vina's relatively simple scoring function, RAPID-Net-guided docking outperforms the state-of-the-art blind-docking tool DiffBindFR~\cite{zhu2024diffbindfr} by a substantial margin (\textbf{54.9\%} vs \textbf{49.1\%}) on the PoseBusters benchmark~\cite{Buttenschoen2024} (see Section~\ref{Evaluation_on_PoseBusters}). This result underscores our central claim: precise pocket identification is a decisive driver of downstream docking success. It is important to note that, unlike DiffBindFR, which incorporates the ligand-induced protein changes, in our approach, we use our pocket predictor to guide AutoDock Vina that docks to a rigid receptor, and all evaluations are performed on holo structures. Nevertheless, the same pocket and search grid localization can boost flexible-receptor engines as well, for example, DSDPFlex~\cite{Dong2024DSDPFlex}:``By leveraging prior knowledge or information, this innovative approach is anticipated to enhance the search process within the appropriate binding pocket.''. Overall, our results indicate that accurate pocket localization currently outweighs receptor flexibility as the key bottleneck, making RAPID-Net a practical route to higher accuracy in binding-site-agnostic workflows and a solid foundation on which ligand-induced conformational refinements can later be added as a subleading effect. We refer the treatment of ligand-induced conformational changes to our future work, focusing here on maximizing the pocket prediction accuracy.

For comparison with co-folding approaches such as AlphaFold 3 (AF3)~\cite{abramson2024accurate}, we follow the PoseBench protocol~\cite{morehead2024posebench}. This protocol uses a curated subset of 130 protein-ligand complexes from PoseBusters~\cite{Buttenschoen2024} deposited after September 30, 2021, specifically designed to evaluate generalization performance rather than memorization~\cite{morehead2024posebench}. On this challenging benchmark, RAPID‑Net\,+\,Vina achieves a Top-1 success rate of \textbf{53.1\%}, only six percentage points below AF3's \textbf{59.5\%}  (see Section~\ref{Evaluation_on_PoseBusters}), while requiring substantially fewer computational resources and offering significantly better scalability for virtual screening.

RAPID-Net supplies an accurate search box for subsequent lightweight rigid-receptor redocking, whereas AF3 predicts the full protein-ligand complex from the protein sequence and the ligand SMILES. The significantly lower computational demand of our approach is exemplified by the 8F4J structure~\cite{Buttenschoen2024}, which AF3 was unable to process as a whole: \textit{``Another PDB entry (8F4J) was too large to inference the entire system (over 5,120 tokens), so we included only protein chains within $20\ \rm{\AA}$ of the ligand of interest.''}. In contrast, RAPID-Net guided Vina to a pose with $\mathrm{RMSD}<2\,\text{\AA}$\footnote{A demonstration video is available at: \url{https://youtu.be/EkUKmoW11pE?si=3aKBkL3ZRq8ibWqo}}, as illustrated in Fig.~\ref{8F4J_PHO_combined_Fig}. Thus, RAPID-Net provides near-state-of-the-art accuracy while remaining practical for large-scale virtual screening.

In terms of the underlying pocket prediction function, existing pocket predicting algorithms can be broadly classified into classical and Machine Learning (ML)-driven approaches~\cite{utges2024comparative}. Classical methods rely on expert-defined rules or heuristics to detect pockets, whereas ML-driven algorithms learn to extract features from protein data without explicit human instructions. In this work, we adopt the definition of Machine Learning by François Chollet as ``the effort to automate intellectual tasks normally performed by humans''~\cite{chollet2021deep}. From this perspective, the main objective of an ML framework is to unveil a meaningful data representation that allows pocket-prediction rules to emerge automatically, rather than being manually hardcoded.

Furthermore, ``Deep Learning (DL) is a specific subfield of Machine Learning: a new take on learning representations from data that puts an emphasis on learning successive layers of increasingly meaningful representations''~\cite{chollet2021deep}. In the context of pocket identification in our study, these layers of data representations are realized through Deep Neural  Networks (DNNs). Finally, in contrast to DL, so-called ``shallow'' learning approaches typically employ only one or a small number of consecutive data representation layers~\cite{chollet2021deep}.

In the following Sections, we describe our DL model for pocket identification, named \textbf{RAPID-Net} (ReLU Activated Pocket Identification for Docking). We tested RAPID-Net on various benchmarks to demonstrate its efficiency and analyzed its predictions on several therapeutically important proteins. In Section~\ref{Overview}, we provide a brief overview of existing pocket prediction algorithms. In Section~\ref{Sec_Architecture}, we discuss the architecture and design rationale behind our model. In Section~\ref{Training_pipeline}, we present the RAPID-Net training pipeline, highlighting its key differences from existing approaches, including the use of ReLU activation in the last layer of the network to operate on a ``soft'' rather than binary labels. Using the RAPID-Net model developed in Sections~\ref{Sec_Architecture} and~\ref{Training_pipeline}, in Section~\ref{Docking_protocol} we integrate it into a docking protocol that is subsequently used for docking. In Section~\ref{Benchmark_Evaluation}, we describe the evaluation metrics used to assess the quality of the model's predictions. Sections~\ref{Evaluation_on_PoseBusters} and~\ref{Astex_Evaluation} report docking results on the PoseBusters~\cite{Buttenschoen2024} and Astex Diverse Set~\cite{hartshorn2007diverse} datasets, respectively. For completeness and to provide a direct side-by-side comparison with existing pocket prediction algorithms, Section~\ref{coach420_BU48_evaluate} evaluates the geometric characteristics of predicted pockets on the Coach420~\cite{Roy2012} and BU48~\cite{Huang2006} datasets. To demonstrate the ability of RAPID-Net to identify distant binding sites, Section~\ref{Distant_Sites} analyzes its predictions for several therapeutically important proteins with known distal sites. Finally, in Section~\ref{Discussion}, we reach our conclusions and outline the potential directions for future work.

\section{Overview of the Available Pocket Prediction Algorithms}
\label{Overview}

Pocket prediction methods exploit a variety of different techniques to identify potential binding sites. Geometry-based techniques like Fpocket~\cite{LeGuilloux2009}, Ligsite~\cite{Hendlich1997} and Surfnet~\cite{Laskowski1995} identify
cavities by analyzing the geometry of the molecular surface of a protein and usually rely on the use of a grid, gaps, spheres, or tessellation~\cite{LeGuilloux2009,Hendlich1997,Laskowski1995,Levitt1992,Kleywegt1994,Liang1998,Brady2000,Weisel2007}. Energy-based methods
such as PocketFinder~\cite{An2005} rely on the calculation of interaction
energies between the protein and chemical group
or probe to identify cavities~\cite{An2005,Goodford1985,An2004,Laurie2005,Ghersi2009,Ngan2012}
. 

Conservation-based methods make use of sequence evolutionary conservation information to find patterns in multiple sequence alignments and identify conserved key residues for ligand binding
site identification~\cite{Armon2001,Pupko2002,Xie2012}. 

Template-based methods rely on structural information from homologues and the assumption that structurally conserved proteins might bind ligands at a similar location~\cite{Zvelebil1987,Wass2010,Roy2012,Yang2013,Lee2013,Brylinski2013}. Combined approaches or meta-predictors combine multiple methods, or the use of multiple types of data to infer ligand
binding sites, e.g., geometric features with sequence
conservation~\cite{Huang2006,Glaser2006,Halgren2009,Capra2009,Huang2009,Bray2009,Brylinski2009}.

Although classical geometry-based tools such as Fpocket are fast and widely applicable, they can sometimes merge individual pockets into a single ``blurred'' cavity, for example, as observed in key SARS-CoV-2 targets~\cite{yin2020structural,gervasoni2020comprehensive}. In Section~\ref{Distant_Sites}, we revisit the Nsp12 protein considered in~\cite{yin2020structural,gervasoni2020comprehensive} using the RAPID-Net model we developed to demonstrate its ability to identify functionally significant sites.

\section{Design Rationale behind our Model}
\label{Sec_Architecture}

\begin{figure}[]{}
\includegraphics[width=\linewidth]{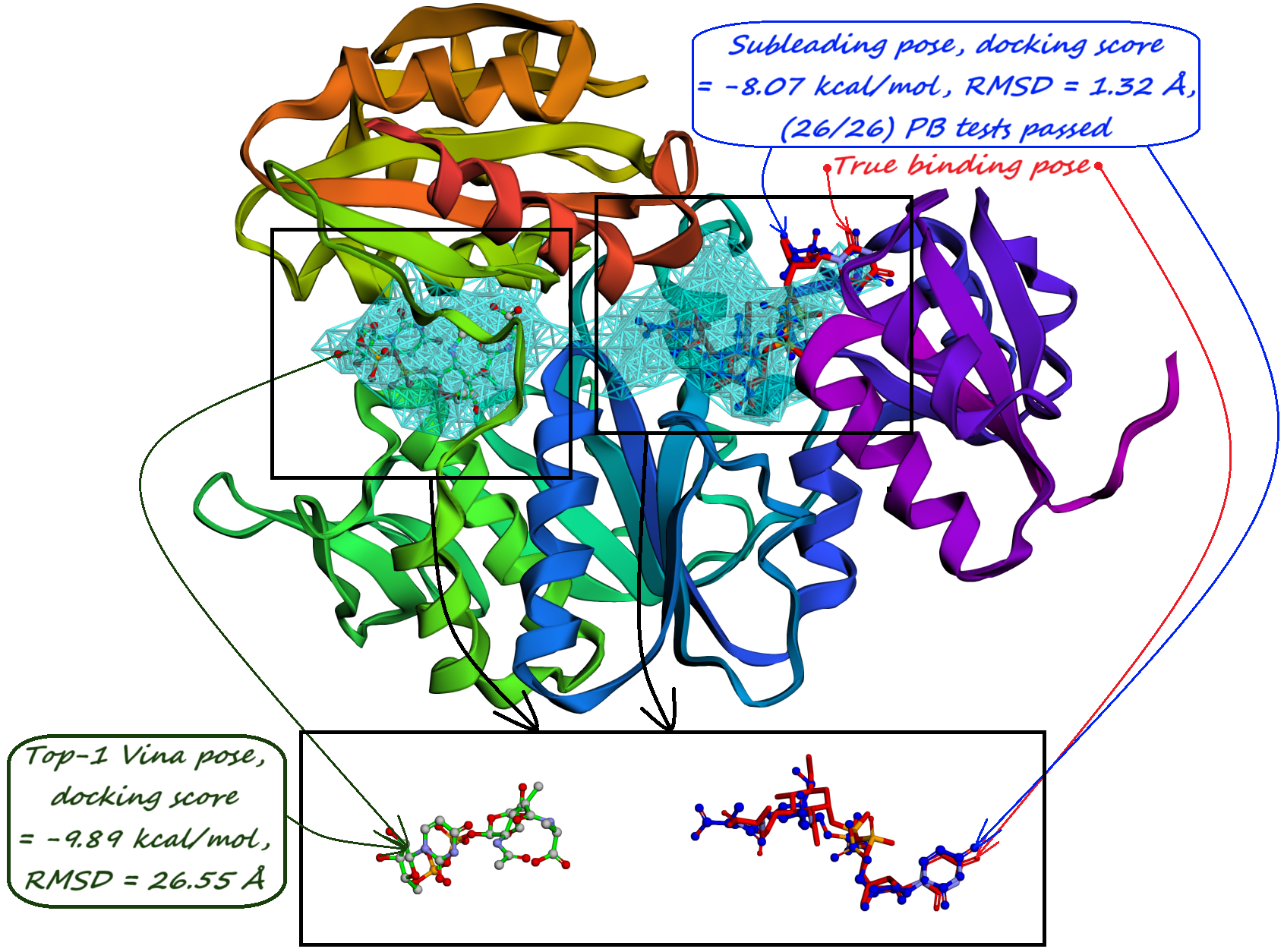}
  \caption{8DP2 protein structure from the PoseBusters~\cite{Buttenschoen2024} dataset. The Top-1 Vina pose is docked in a predicted sub-pocket devoid of true ligand binding pose, while the subleading pose correctly occupies a second predicted sub-pocket and passes all validation tests.}
\label{8DP2_UMA_Fig}
\end{figure}

\begin{figure*}[]{}
  \centering
  \includegraphics[width=\linewidth]{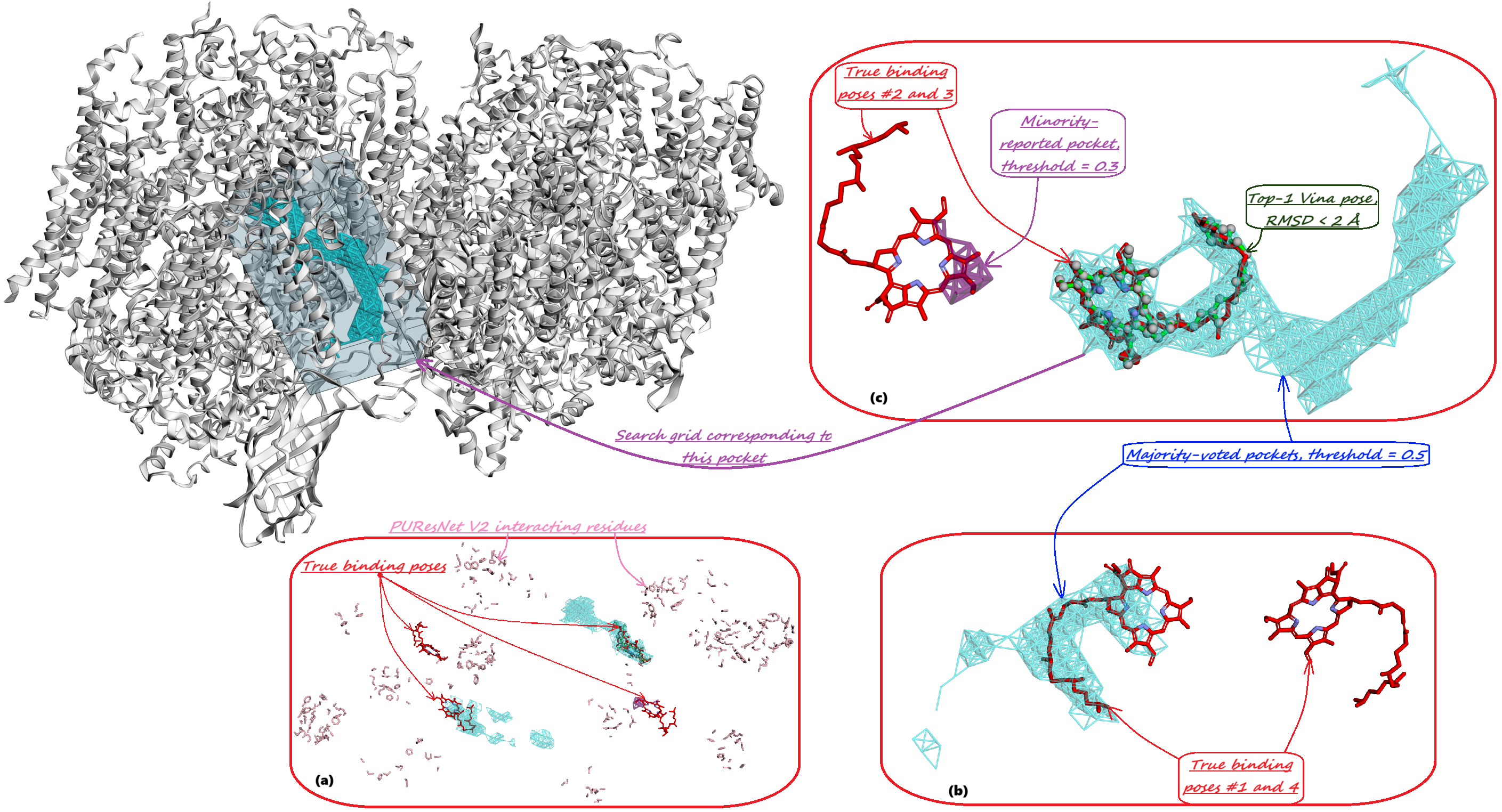}
  \caption{For the 8F4J protein structure from the PoseBusters~\cite{Buttenschoen2024} dataset, our docking approach achieves $\text{RMSD} < 2$\,\AA, while AlphaFold 3~\cite{abramson2024accurate} is unable to process the full structure due to input size limitations. Unlike residue-focused methods such as PUResNet V2~\cite{jeevan2024puresnetv2}, which produce complex predictions that are difficult to interpret, our approach offers clear, plug-and-play guidance for docking grid selection, as illustrated in the bottom panel.}

  \label{8F4J_PHO_combined_Fig}
\end{figure*}

\begin{figure*}[]{}
  \centering
  \includegraphics[width=\linewidth]{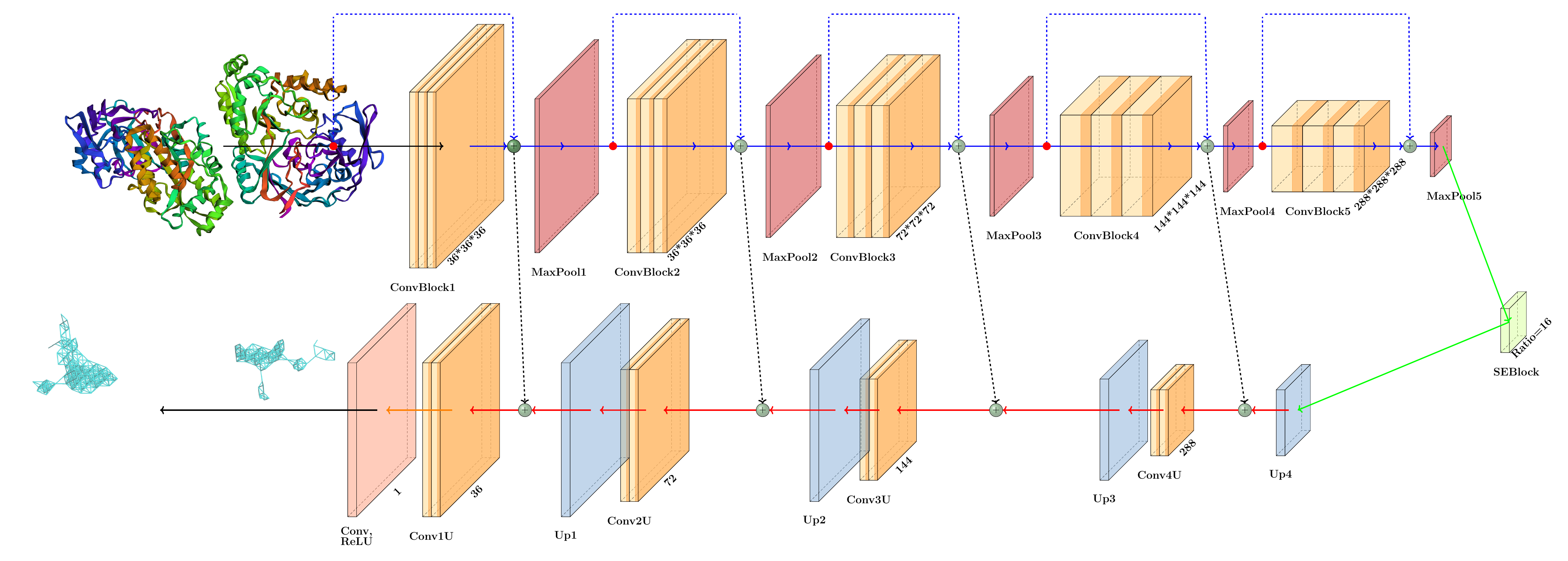}
  \caption{Schematic representation of \textbf{RAPID-Net} (ReLU-Activated Pocket Identification for Docking). Key improvements that distinguish RAPID-Net from previous approaches include a ReLU activation in the final layer, the usage of a soft dice loss function, including a single SE-attention block, and removing redundant residual connections.}
  \label{RAPID_Fig}
\end{figure*}

When predicting likely binding pockets, there is a well-known trade-off between precision and recall. A recent evaluation of pocket prediction methods based on the geometrical properties of their predicted pockets~\cite{utges2024comparative} highlights this trade-off. Currently available ML methods such as VN-EGNN~\cite{sestak2024vn_egnn}, GrASP~\cite{smith2024graph}, and PUResNet~\cite{kandel2021puresnet} achieve high precision (over 90\%), but they systematically predict a small number of pockets, leading to low recall. As~\cite{utges2024comparative} points out, generating multiple predictions, some of which may be false positives, is often more useful than potentially missing viable binding sites.

To address this shortcoming and mitigate low recall, we develop RAPID-Net as an ensemble-based model to improve the stability of prediction accuracy and coverage of potential binding sites. RAPID-Net consists of five independently trained model replicas to aggregate results and increase the reliability of its predictions. For subsequent docking, RAPID-Net returns two types of pockets:

\begin{enumerate}
    \item \textit{Majority-voted pockets}: Consisting of voxels predicted by at least 3 out of 5 ensemble models, for high-confidence predictions.
    
    \item \textit{Minority-reported pockets}: Consisting of voxels predicted by at least 1 out of 5 ensemble models, increasing the overall recall.
\end{enumerate}

As demonstrated on diverse benchmark datasets in the following Sections, including PoseBusters~\cite{Buttenschoen2024}, Astex Diverse Set~\cite{hartshorn2007diverse}, Coach420~\cite{Roy2012}, and BU48~\cite{Huang2006}, the proposed ensemble-based approach yields more stable and reliable performance compared to current ML-driven pocket predictors, both considering the majority-voted and minority-reported pockets.

For instance, consider the 8DP2 protein from the PoseBusters~\cite{Buttenschoen2024} dataset, illustrated in Fig.~\ref{8DP2_UMA_Fig}. RAPID-Net predicts two majority-voted sub-pockets connected by a link. However, because no ligand binding occurs in one of these sub-pockets, it constitutes a false positive. By contrast, one of the subleading poses binds within the second predicted sub-pocket and passes all tests. This example demonstrates RAPID-Net's ability to balance accuracy and recall, providing a comprehensive and robust pocket prediction framework for complex docking tasks. Furthermore, as shown in the following Sections, the minority-reported pockets predicted by our model are often shallow pockets corresponding to secondary binding sites, in contrast to the more prominent majority-voted pockets.

\vspace{1em}
\noindent
\textbf{Types of outputs from pocket predictors.}
The type of output produced by pocket predictors can be divided into three main categories. 

Kalasanty~\cite{stepniewska2020improving} and PUResNet V1\footnote{PUResNet V1\cite{kandel2021puresnet} predicts cavities where the ligand is likely to reside, whereas PUResNet V2~\cite{jeevan2024puresnetv2} predicts residues likely to interact with the ligand. For simplicity, we refer to PUResNet~V1 as ``PUResNet'' throughout the text, as in this work we focus on cavities for generating an accurate search grid for subsequent docking.}~\cite{kandel2021puresnet} both detect potential ligand-binding pockets using a voxel-based representation, where each voxel is $2\text{\AA} \times 2\text{\AA} \times 2\text{\AA}$ in size.

In contrast, PRANK~\cite{krivak2015improving} and DeepPocket~\cite{aggarwal2021deeppocket} reweight the alpha-spheres identified by the rule-based FPocket~\cite{le2009fpocket} algorithm. As a result, these methods inherit FPocket's initial search space and any associated cascading errors. Moreover, alpha-sphere-based approaches may lack sufficient granularity and can struggle to capture subtle variations among binding sites, as previously noted~\cite{yin2020structural,gervasoni2020comprehensive}.

GrASP~\cite{smith2024graph}, IF-SitePred~\cite{carbery2024learnt}, VN-EGNN~\cite{sestak2024vn_egnn}, and PUResNetV2~\cite{jeevan2024puresnetv2} provide predictions of potential ligand interactions at the residue level. Although these predictions can reveal important binding details, they are less suitable for docking workflows that require a well-defined three-dimensional region for accurate ligand placement and orientation. By precisely defining these regions, the computational search space is reduced as docking is restricted to reasonable binding pockets. As illustrated in the bottom panel of Fig.~\ref{8F4J_PHO_combined_Fig}, residue-level predictions can be difficult to interpret when defining a reasonable search grid.

Although some studies reweight residues to guide blind docking~\cite{shen2024pgbind}, and others~\cite{huang2023dsdp} employ voxelized pocket prediction models--such as PUResNet \cite{kandel2021puresnet}--in tandem with AutoDock Vina \cite{eberhardt2021autodock}, our work focuses on developing a voxel-based cavity prediction model that seamlessly integrates into the docking pipeline. This voxel-based approach not only provides well-defined search regions but also facilitates a more modular and interpretable workflow compared to less structured docking strategies, as illustrated in Fig.~\ref{Vina_Setup_Fig} and discussed further in Section~\ref{Docking_protocol}. 

Furthermore, as discussed in Section~\ref{Evaluation_on_PoseBusters}, improving the accuracy of pocket identification directly improves docking results. In particular, AutoDock Vina~\cite{eberhardt2021autodock}, guided by our pocket predictor, outperforms the state-of-the-art DiffBindFR~\cite{zhu2024diffbindfr} docking tool, which otherwise scans the entire protein surface in ``blind'' settings.

\vspace{1em}
\noindent
\textbf{Our model architecture.}
Fig.~\ref{RAPID_Fig} depicts the architecture of our proposed RAPID-Net model, which is similar to U-Net~\cite{ronneberger2015u} with encoder and decoder branches. However, we implement several notable adaptations. 

First, we include residual connections \textit{only} in the encoder part of the model, as our experiments show that they are highly beneficial there, but lead to overfitting if included in the decoder. This approach differs from Kalasanty~\cite{stepniewska2020improving}, which omits residual blocks altogether~\cite{he2016deep}, and PUResNet~\cite{kandel2021puresnet}, which uses them throughout the network.

Second, although the standard SE-ResNet~\cite{hu2018squeeze} architecture typically includes attention blocks at multiple layers, studies on breast-cancer imaging~\cite{jiang2019breast} and Raman spectra classification~\cite{balytskyi2024enhancing} have shown that using a \textit{single} attention block can be highly effective while mitigating overfitting. Our experiments similarly suggest that in inherently noisy datasets, adding too many attention modules can amplify noise and degrade performance.

As we demonstrate in the following Sections, adopting a moderate level of attention, limiting residual connections, and incorporating a modified loss function significantly improves performance compared to earlier models.

\section{Model Training and Inference Pipeline}
\label{Training_pipeline}

We trained our model using the sc-PDB~\cite{desaphy2015scpdb} dataset, which contains protein structures with annotated binding sites. In this dataset, cavities are defined using VolSite~\cite{desaphy2012comparison}, which maps the pharmacophore properties of nearby protein atoms onto a three-dimensional grid. This method assumes that ligand-induced conformational changes remain relatively small~\cite{kellenberger2008similarity}, treating each cavity atom as a ``pseudoatom'' to denote an interaction point rather than a physical atom.

Following~\cite{kandel2021puresnet}, we used a curated subset of sc-PDB~\cite{desaphy2015scpdb} in which redundant protein structures were filtered out based on their Tanimoto index~\cite{bajusz2015tanimoto}. Each protein structure was then placed in a $36 \times 36 \times 36$ grid, with each voxel corresponding to a $2\text{\AA} \times 2\text{\AA} \times 2\text{\AA}$ unit cell. We extracted 18 features per voxel using the \texttt{tfbio}~\cite{tfbio2018} package to provide a representation of the protein environment for the model training.

In contrast to previous studies, we apply the medical image segmentation practices~\cite{gros2021softseg,wang2023dice1,milletari2016vnet} to perform \textit{soft} segmentation of ligand-binding pockets, rather than treating it as a strict binary classification problem. In Kalasanty~\cite{stepniewska2020improving} and PUResNet~\cite{kandel2021puresnet}, each voxel is considered part of a binding pocket if it contains at least one cavity pseudoatom. The number of pseudoatoms per voxel is then clipped to the range $[0,1]$, and the model is trained with a Dice loss function as part of a binary classification task:

\begin{equation}
\Delta_{\text{Dice}}\left(V^{\text{clipped}}_{\text{true}}, V^{\text{clipped}}_{\text{predicted}}\right)
= 1
- \frac{2 \left\lvert V^{\text{clipped}}_{\text{true}} \cap V^{\text{clipped}}_{\text{predicted}} \right\rvert}
       {\left\lvert V^{\text{clipped}}_{\text{true}} \right\rvert 
        + \left\lvert V^{\text{clipped}}_{\text{predicted}} \right\rvert},
\label{Dice_Eqn}
\end{equation}
where $V^{\text{clipped}}_{\text{true}}$ and $V^{\text{clipped}}_{\text{predicted}}$ represent the binary masks of the true and predicted pockets, respectively, $\cap$ indicates their intersection, and $\left\lvert V^{\text{clipped}}_{\text{true}} \right\rvert$ and $\left\lvert V^{\text{clipped}}_{\text{predicted}} \right\rvert$ denote the number of occupied voxels in the true and predicted masks, respectively.

Recently, PUResNet V2~\cite{jeevan2024puresnetv2} sought to improve performance by replacing the Dice loss function with a focal loss~\cite{lin2017focal}. However, our extensive experiments indicate that adopting a ``soft'' approach~\cite{gros2021softseg,wang2023dice1,milletari2016vnet} is better suited in this context. In contrast to the binary ``yes/no'' classification of occupied voxels, the density of pseudoatoms serves as an indicator of the proximity of a voxel to the interior or boundary of a pocket. Higher densities typically occur near polar or charged residues and ligand functional groups, whereas lower densities are often associated with hydrophobic regions.

Drawing inspiration from medical image segmentation methods~\cite{gros2021softseg,wang2023dice1,milletari2016vnet}, we replace the sigmoid output in the final layer with a ReLU~\cite{nair2010relu} (hence the name of our model, ReLU Activated Pocket Identification for Docking, \textbf{RAPID-Net}):
\begin{equation}
\sigma\left(x\right) = \frac{1}{1 + e^{-x}} 
\;\;\rightarrow\;\;
\text{ReLU}\left(x\right) = \max\left(0, x\right)
\end{equation}

For the model training, we also replace the conventional Dice loss in Eqn.~\ref{Dice_Eqn} with its ``soft'' variant:
\begin{equation}
\Delta_{\text{Soft Dice}, \ \text{L}^2}(V_{\text{true}}, V_{\text{predicted}}) 
= 1
- \frac{2 \sum_{\text{Grid}} \left(V_{\text{true}} \, V_{\text{predicted}}\right)}
       {\sum_{\text{Grid}} \left(V_{\text{true}}^2 
        +  V_{\text{predicted}}^2\right)}
\label{Soft_Dice_Eqn}
\end{equation}
Several ways of implementing the ``soft'' Dice loss have been proposed~\cite{wang2023dice1,milletari2016vnet}, but we found that this simplest version, based on the $\text{L}^2$ norm, works best.

To mitigate class imbalance, as the number of non-interacting voxels sharply exceeds the number of interacting ones, we applied class reweighting using \texttt{scikit-learn}~\cite{scikit-learn}. During inference, a voxel is classified as part of a binding pocket if the model output exceeds $\texttt{threshold}\,=0.5$. When ensembling five models, we apply morphological closing using the \texttt{binary\_closing} function from the \texttt{scipy.ndimage} package~\cite{2020SciPy-NMeth} to mitigate sparsity in the predicted pocket regions across model replicas.

Unlike previous approaches that use \texttt{cavity6.mol2} labels~\cite{kandel2021puresnet,stepniewska2020improving}, we train our model with threshold-less pocket labels (\texttt{cavityALL.mol2}). In \texttt{cavity6.mol2}, annotations are restricted to regions within $6.5\,\text{\AA}$ of the ligand’s heavy atoms~\cite{desaphy2015scpdb}, potentially overlooking distal functional regions such as allosteric pockets, exosites, or flexible loops. These sites can critically influence drug binding and resistance, making them key therapeutic targets. Secondary binding sites, which often lie beyond the $6.5\,\text{\AA}$ boundary, mediate interactions with larger substrates or cofactors and help position them within the catalytic pocket, as further discussed in Section~\ref{Distant_Sites}. By adopting threshold-less labels, our model identifies both catalytic and more distant sites, thereby improving predictions for proteins that contain such secondary binding regions.

This approach also makes the Distance Center Center (DCC) metric, which was previously used to evaluate Kalasanty~\cite{stepniewska2020improving} and PUResNet~\cite{kandel2021puresnet}, less relevant. DCC defines the distance between the geometrical center of the predicted pocket and the true ligand binding pose, but ``tunnels'' can extend from the primary binding pocket to distal residues that lie far from the ligand but have significant therapeutic implications (see Section~\ref{Distant_Sites} for examples). Furthermore, the threshold-less pocket definition is less sensitive to the ligand size, thereby improving the generalizability of our model.

\section{Pocket-informed Docking Protocol and its Rationale}
\label{Docking_protocol}
\begin{figure}[]{}
\includegraphics[width=\linewidth]{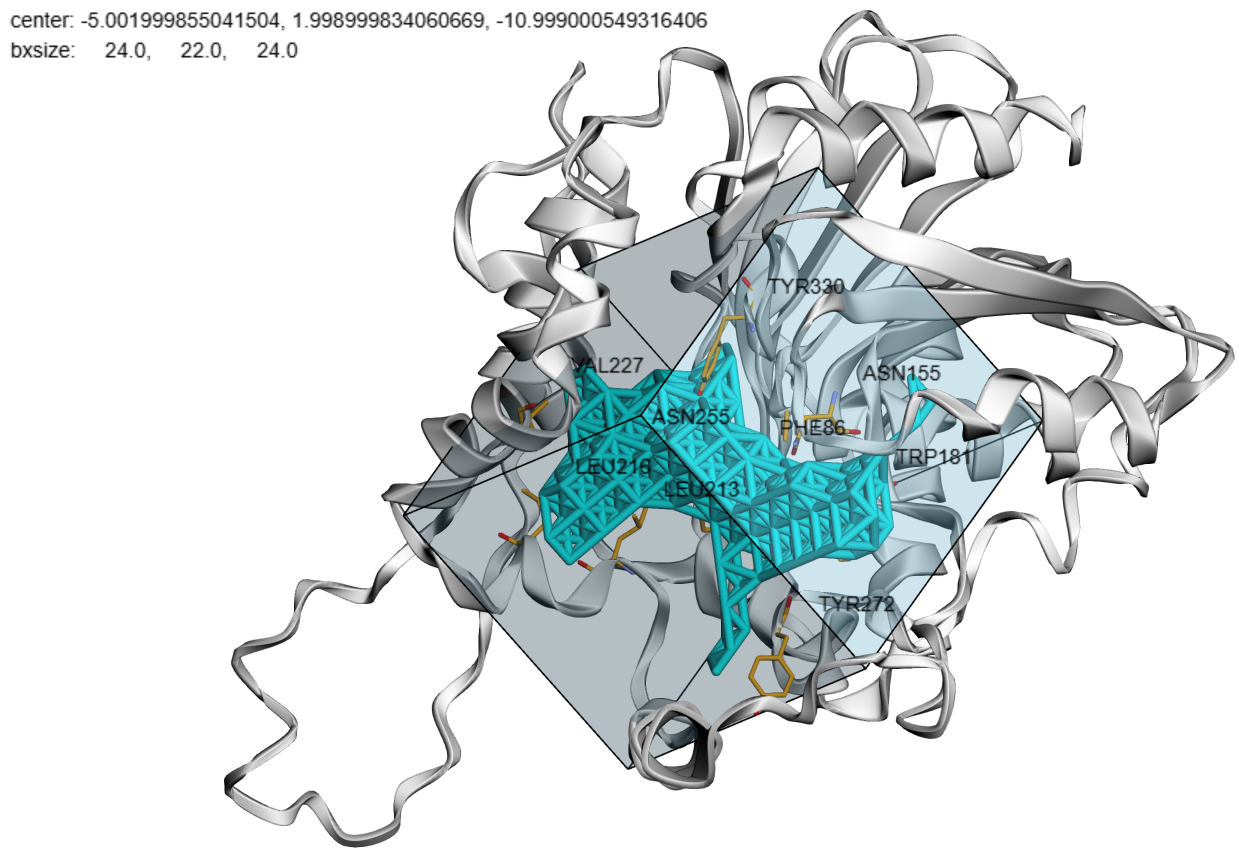}
  \caption{Majority-voted pocket predicted by RAPID-Net and its corresponding search grid with $2$ $\text{\AA}$ threshold for ABHD5 protein~\cite{varadi2022alphafold}.}
\label{Vina_Setup_Fig}
\end{figure}

For each predicted pocket, we define the center of the search grid as the average of the maximum and minimum $x$, $y$, and $z$ coordinates of all pseudoatoms in the pocket:

\begin{equation}
\text{center}\left(x,y,z\right) = \frac{\texttt{max}\left(x,y,z\right) + \texttt{min}\left(x,y,z\right)}{2},
\end{equation}
where $\texttt{max}\left(x,y,z\right)$ and $\texttt{min}\left(x,y,z\right)$  are the maximum and minimum coordinates, respectively. The grid dimensions are then given by:
\begin{equation}
\Delta x,\Delta y,\Delta z = (\texttt{max}\left(x,y,z\right) - \texttt{min}\left(x,y,z\right)) + 2 \cdot \text{threshold}
\end{equation}
Fig.~\ref{Vina_Setup_Fig} illustrates this setup for the majority-voted pocket in the ABHD5 protein~\cite{varadi2022alphafold} with a threshold value of $2$ $\text{\AA}$.

\begin{figure}[]{}
\includegraphics[width=\linewidth]{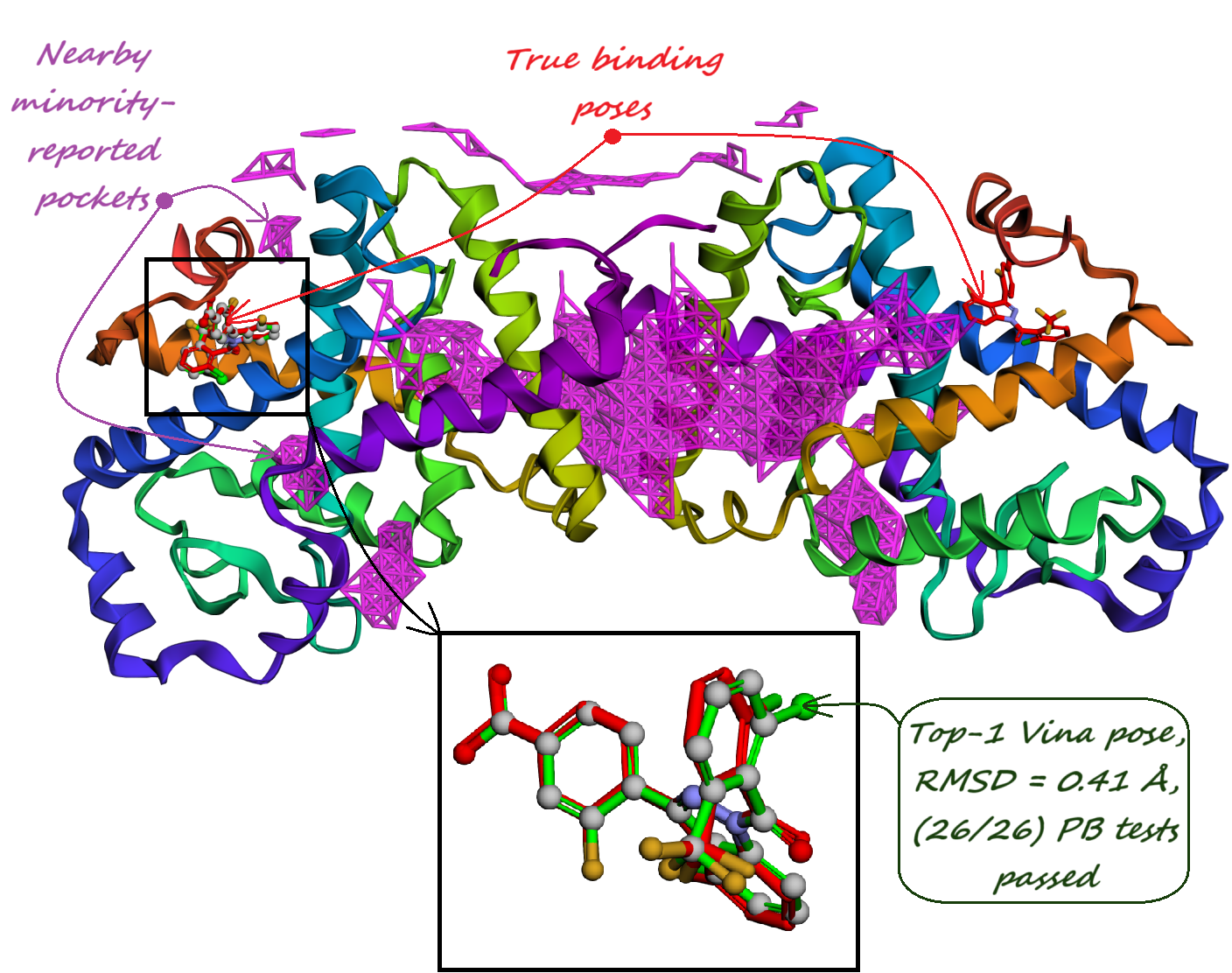}
  \caption{For the 8FAV protein structure, our model predictions are inaccurate, and none of the true ligand binding poses are within one of the predicted pockets. Nevertheless, the docking is successful because the search grids with large thresholds smooths out this drawback.}
\label{8FAV_4Y5_Fig}
\end{figure}

For each majority-voted pocket, we generate search grids with thresholds of $2\,\text{\AA}$ and $5\,\text{\AA}$. For minority-reported pockets, we use thresholds of $2\,\text{\AA}$, $5\,\text{\AA}$, $10\,\text{\AA}$, and $15\,\text{\AA}$. This strategy accounts for large ligands, such as those in the PoseBusters~\cite{Buttenschoen2024} dataset, which may extend beyond the predicted pockets, and also reduces potential inaccuracies in our model's pocket predictions.

As an example, Fig.~\ref{8FAV_4Y5_Fig} shows the 8FAV protein from the PoseBusters dataset~\cite{Buttenschoen2024}, where none of the predicted pockets overlap with one of the true ligand binding poses. However, the docking still succeeds because the expanded search grid thresholds encompass the true binding site. This case demonstrates how geometry-based evaluation metrics, such as those in~\cite{utges2024comparative}, can overlook the practical success of docking, even if the pocket predictions are imperfect.

To demonstrate that the accuracy of our model is primarily a result of architectural and training improvements--not merely ensembling--we also provide single-model predictions from Kalasanty~\cite{stepniewska2020improving}, PUResNet~\cite{kandel2021puresnet}, and individual RAPID-Net runs. Similarly, for each predicted pocket, we generate grids with $2\,\text{\AA}$, $5\,\text{\AA}$, $10\,\text{\AA}$, and $15\,\text{\AA}$ thresholds.

Finally, we perform molecular docking on each identified grid using AutoDock Vina 1.2.5~\cite{eberhardt2021autodock}, with default settings (\texttt{Exhaustiveness}$ = 32$, \texttt{num\_modes}$ = 40$), following the protocol described in~\cite{Buttenschoen2024}.

\section{Test Benchmarks and Evaluation Metrics}
\label{Benchmark_Evaluation}

We evaluate docking performance as the percentage of Top-1 Vina poses with a root mean square deviation (RMSD) below $2$ $\text{\AA}$ between the predicted and one of the true ligand poses if multiple true poses are available. RMSD is calculated as:

\begin{equation}
\text{RMSD} = \sqrt{\frac{1}{N} \sum_{i=1}^{N} \lVert \mathbf{r}_i^{\text{true}} - \mathbf{r}_i^{\text{pred}} \rVert^2},
\end{equation}
where $N$ is the number of ligand atoms, $\mathbf{r}_i^{\text{true}}$ and $\mathbf{r}_i^{\text{pred}}$ are positions of the $i$-th atom in the true and predicted poses, respectively, and $\lVert \cdot \rVert$ is the Euclidean norm. RMSD is computed using the \texttt{CalcRMS} function in \texttt{RDKit}~\cite{landrum2013rdkit}.

Docking methods like AutoDock Vina~\cite{eberhardt2021autodock} balance speed and accuracy~\cite{su2018comparative}, utilizing ``fast and dirty'' scoring functions. A common strategy to enhance their accuracy is to generate an ensemble of potential poses and then identify the best pose among them using more accurate scoring tools, for example~\cite{li2017delphiscorer,shen2022boosting}. Fast scoring functions enable the efficient exploration of numerous poses, while more accurate methods are reserved for refinement and final pose selection. To account for this approach, we also report the sampling RMSD accuracy, defined as the percentage of cases with ``at least one correct pose in the ensemble''. This metric underscores the potential gains that can be achieved through more accurate rescoring. Since our primary focus is on improving pocket identification rather than developing new rescoring functions, we defer such work to the future. Additionally, we evaluate the PB-success rate, defined as the percentage of poses with $\text{RMSD} < 2$ \AA\ that pass all chemical validity checks defined by PoseBusters~\cite{Buttenschoen2024}.

Finally, similarly to~\cite{kandel2021puresnet}, we compute the Pocket-Ligand Intersection (PLI) score to quantify the proportion of ligand atoms residing within the predicted pocket. Unlike~\cite{kandel2021puresnet}, which computes a voxel-based intersection, we measure the average fraction of ligand heavy atoms located within $5$ \AA\ of at least one pocket pseudoatom, yielding a more intuitive metric:

\begin{equation}
\text{PLI} = \frac{N_{\text{within 5 \AA}}}{N_{\text{total}}},
\end{equation}
where $N_{\text{within 5 \AA}}$ is the number of ligand heavy atoms within $5$ \AA\ of any pocket pseudoatom, and $N_{\text{total}}$ is the total number of heavy atoms in the ligand.

However, unlike~\cite{kandel2021puresnet}, we compute this ratio for \textit{all} protein-ligand pairs in the dataset rather than restricting it to cases where the ligand is centered within $\le 4\,\text{\AA}$ of the pocket center. If no pockets are predicted, the PLI is set to zero. When a protein has multiple predicted pockets or multiple ligand poses, we report the maximum PLI value across all possibilities. For completeness, we also indicate how many proteins have at least one predicted pocket and how many have at least one ligand pose within the defined search grids.

Our primary dataset for evaluating docking performance is the PoseBusters~\cite{Buttenschoen2024} dataset, which consists of novel protein-ligand complexes that were not available during the development of current docking programs and thus provides a rigorous assessment of their generalization capabilities. For instance, several docking algorithms that claimed superior accuracy on CASF 2016~\cite{su2018comparative} dataset, perform drastically worse on PoseBusters~\cite{Buttenschoen2024}. Additionally, we evaluate docking accuracy on the Astex Diverse Set~\cite{hartshorn2007diverse}. Although this dataset is older than PoseBusters~\cite{Buttenschoen2024}, it remains a challenging benchmark for many algorithms that had previously reported high accuracy on CASF 2016~\cite{su2018comparative}.

We distinguish between two docking scenarios: those that use ``prior knowledge'' of the binding site (which is typically unavailable in real-world applications) and truly ``blind'' settings that lack this information. In both the PoseBusters~\cite{Buttenschoen2024} and Astex~\cite{hartshorn2007diverse} datasets, when the binding site is assumed to be known, the reference ligand coordinates are used to center a $25\,\text{\AA} \times 25\,\text{\AA} \times 25\,\text{\AA}$ bounding box. However, to evaluate RAPID-Net under truly ``blind'', binding-site-agnostic conditions, we omit these reference ligands entirely.

In addition to docking accuracy, for completeness, we report the PLI rates for PoseBusters~\cite{Buttenschoen2024} and Astex~\cite{hartshorn2007diverse} datasets. For direct comparison with PUResNet~\cite{kandel2021puresnet}, we also evaluate PLI for the Coach420~\cite{Roy2012} and BU48~\cite{Huang2006} datasets, excluding any structures that appear in the training set as specified in~\cite{kandel2021puresnet}. Lastly, to highlight the advantages of our model architecture and training approach, we report docking accuracy for both single RAPID-Net models and their ensembled version. PLI is calculated for individual models, majority-voted pockets, and minority-reported pockets.

\section{Evaluation on the PoseBusters Dataset}
\label{Evaluation_on_PoseBusters}

\begin{figure}[]{}
\includegraphics[width=\linewidth]{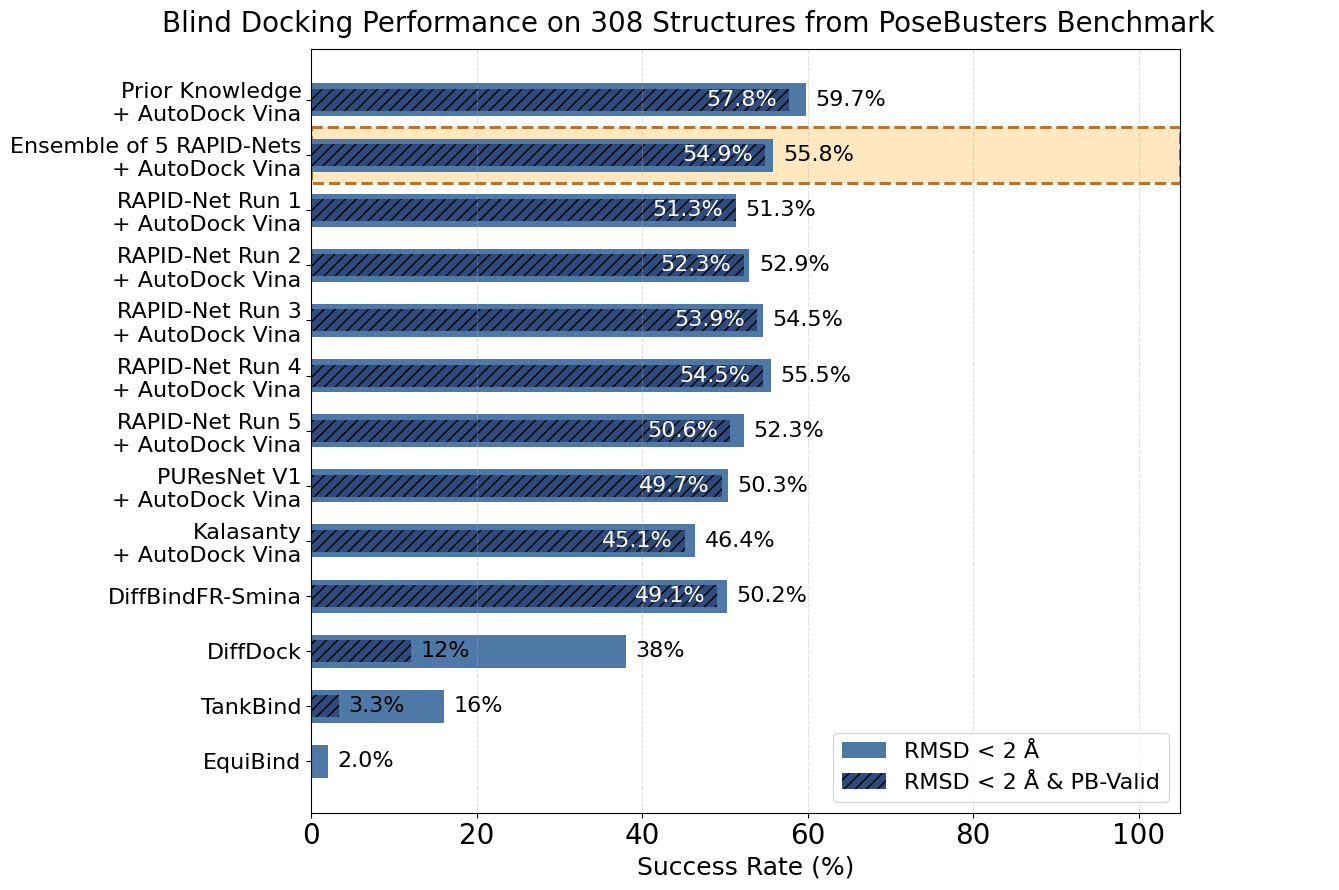}
  \caption{Comparison of Top-1 Vina accuracies when guided by different pocket identification algorithms, and a DiffBindFR-Smina~\cite{zhu2024diffbindfr}, for PoseBusters~\cite{Buttenschoen2024} dataset.}
\label{PB_blind_docking_results_Fig}
\end{figure}

We obtained the following results by docking the PoseBusters~\cite{Buttenschoen2024} dataset using the aforementioned protocol. Fig.~\ref{PB_blind_docking_results_Fig} shows the percentage of predictions with $\text{RMSD} < 2\,\text{\AA}$ and the corresponding PB-success rates. When guided by the ensembled RAPID-Net, AutoDock Vina~\cite{eberhardt2021autodock} achieved \textbf{55.8\%} RMSD-correct predictions and a \textbf{54.9\%} PB-success rate, outperforming DiffBindFR~\cite{zhu2024diffbindfr}, which scored \textbf{50.2\%} and \textbf{49.1\%}, respectively. This result highlights the crucial impact of accurate pocket identification on the subsequent docking accuracy. Our approach uses targeted docking with the widely used AutoDock Vina~\cite{eberhardt2021autodock}, whereas DiffBindFR~\cite{zhu2024diffbindfr} scans the entire protein surface.

\begin{figure}[]{}
\includegraphics[width=\linewidth]{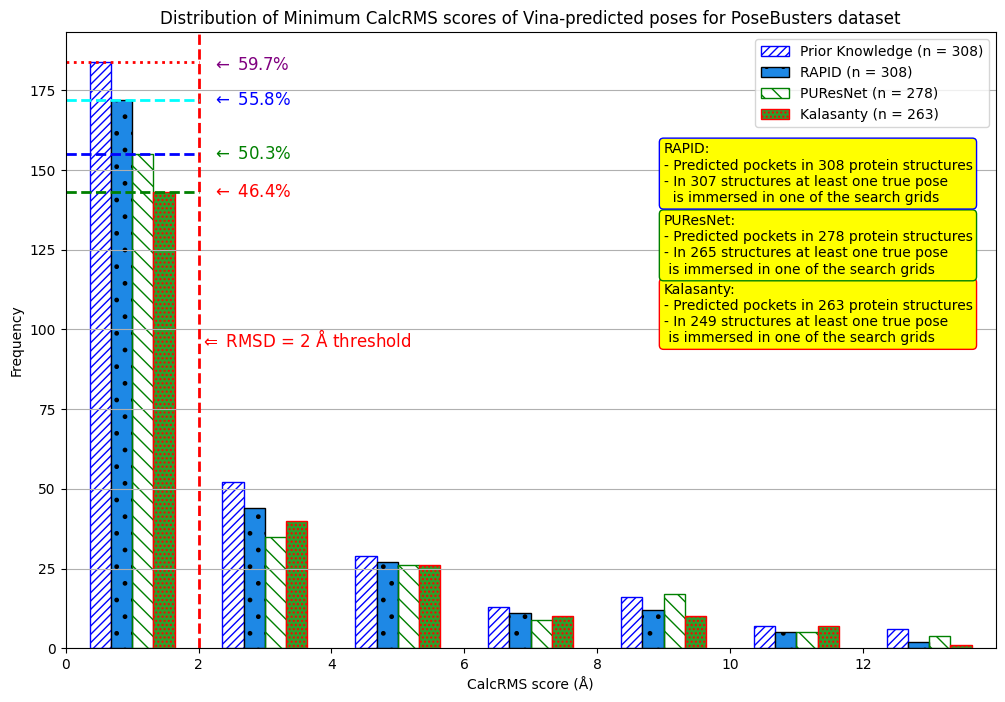}
\caption{Distribution of RMSD values for Top-1 Vina predicted poses in the PoseBusters~\cite{Buttenschoen2024} dataset. When multiple true ligand poses are available, RMSD to the closest one is reported.}
\label{PB_RMSD_Fig}
\end{figure}

\begin{figure}[]{}
\includegraphics[width=\linewidth]{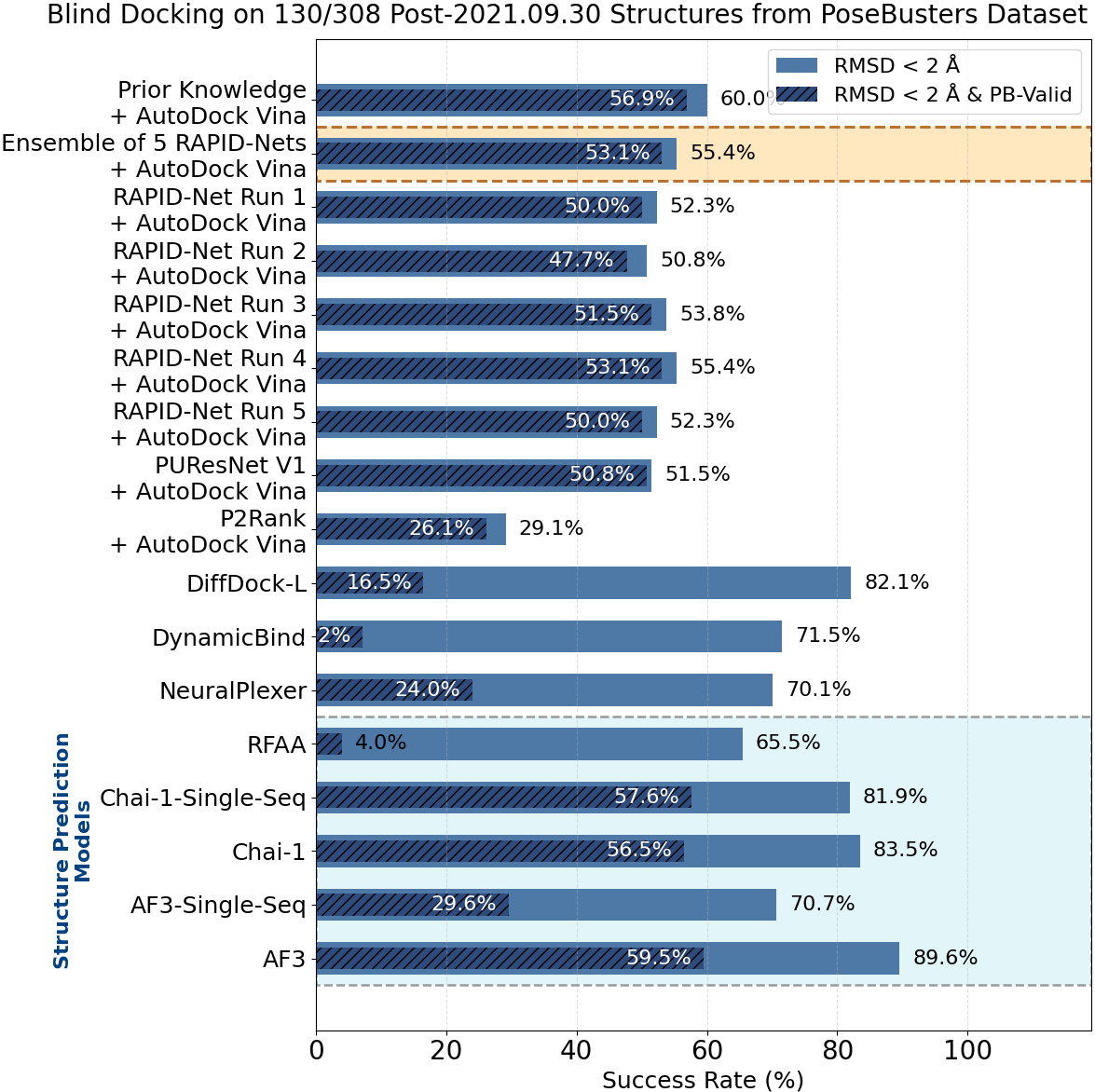}
\caption{Performance comparison on the subset of novel PoseBusters structures as suggested in~\cite{morehead2024posebench}. The RAPID-Net-guided AutoDock Vina achieves \textbf{53.1\%} compared to \textbf{59.5\%} of AlphaFold 3, with significantly lower computational costs.}
\label{PB_130_blind_docking_results_Fig}
\end{figure}

\begin{figure}[]{}
\includegraphics[width=\linewidth]{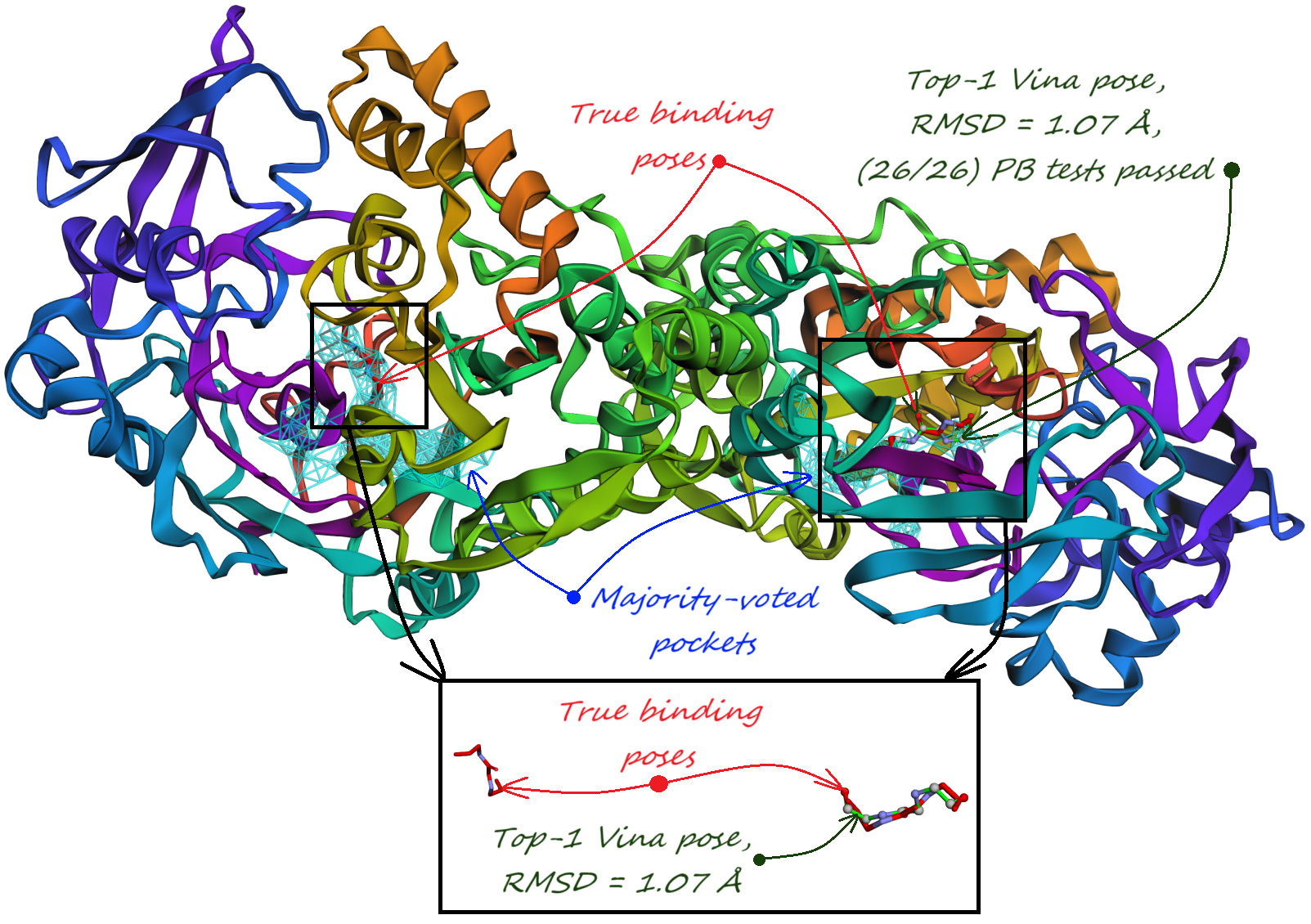}
  \caption{In 7KZ9 protein structure, RAPID-Net predicts majority-voted pockets around both true binding poses, and the Top-1 Vina pose passes all tests.}
\label{7KZ9_XN7_Fig}
\end{figure}

\begin{figure}[]{}
\includegraphics[width=\linewidth]{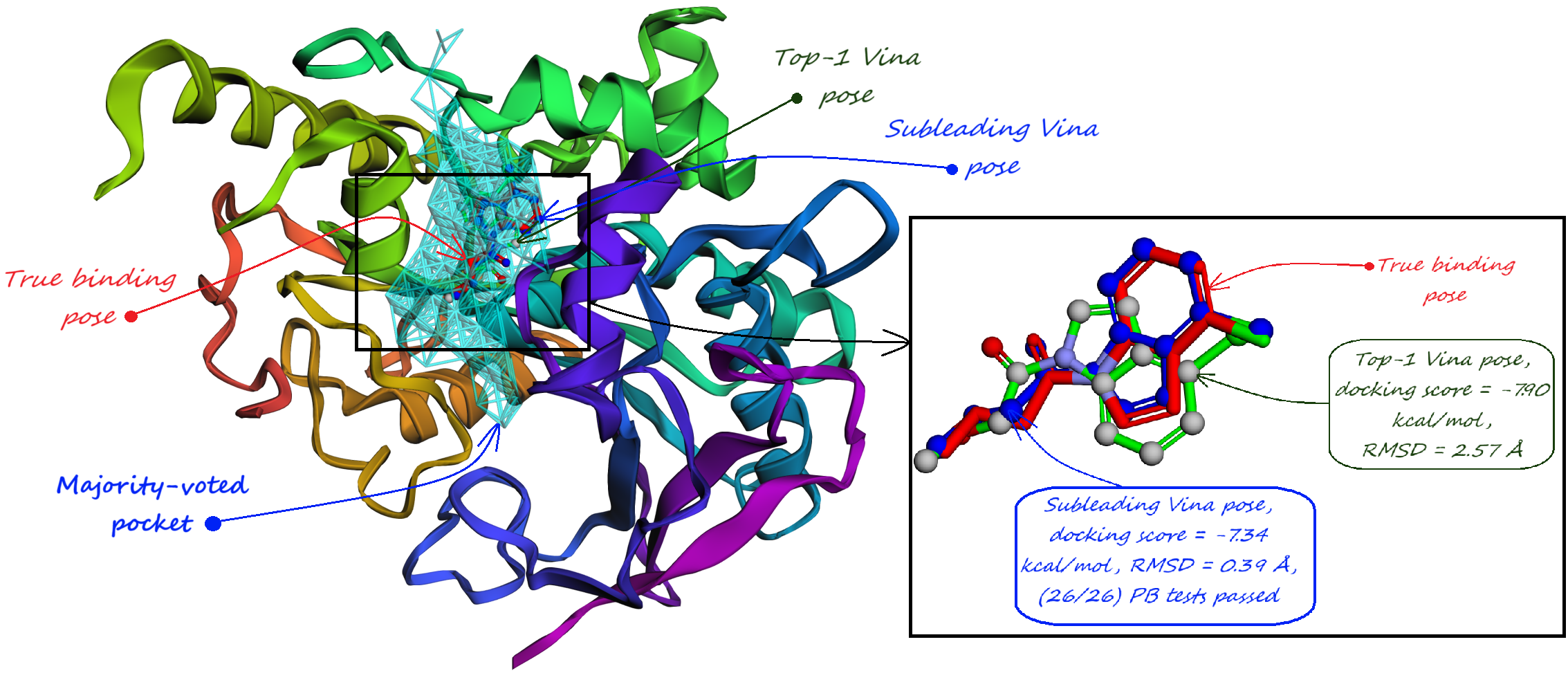}
\caption{For the 8BTI protein structure, the Top-1 Vina pose is incorrect, but the subleading pose passes all tests.}
\label{8BTI_RFO_Fig}
\end{figure}

\begin{figure}[]{}
\includegraphics[width=\linewidth]{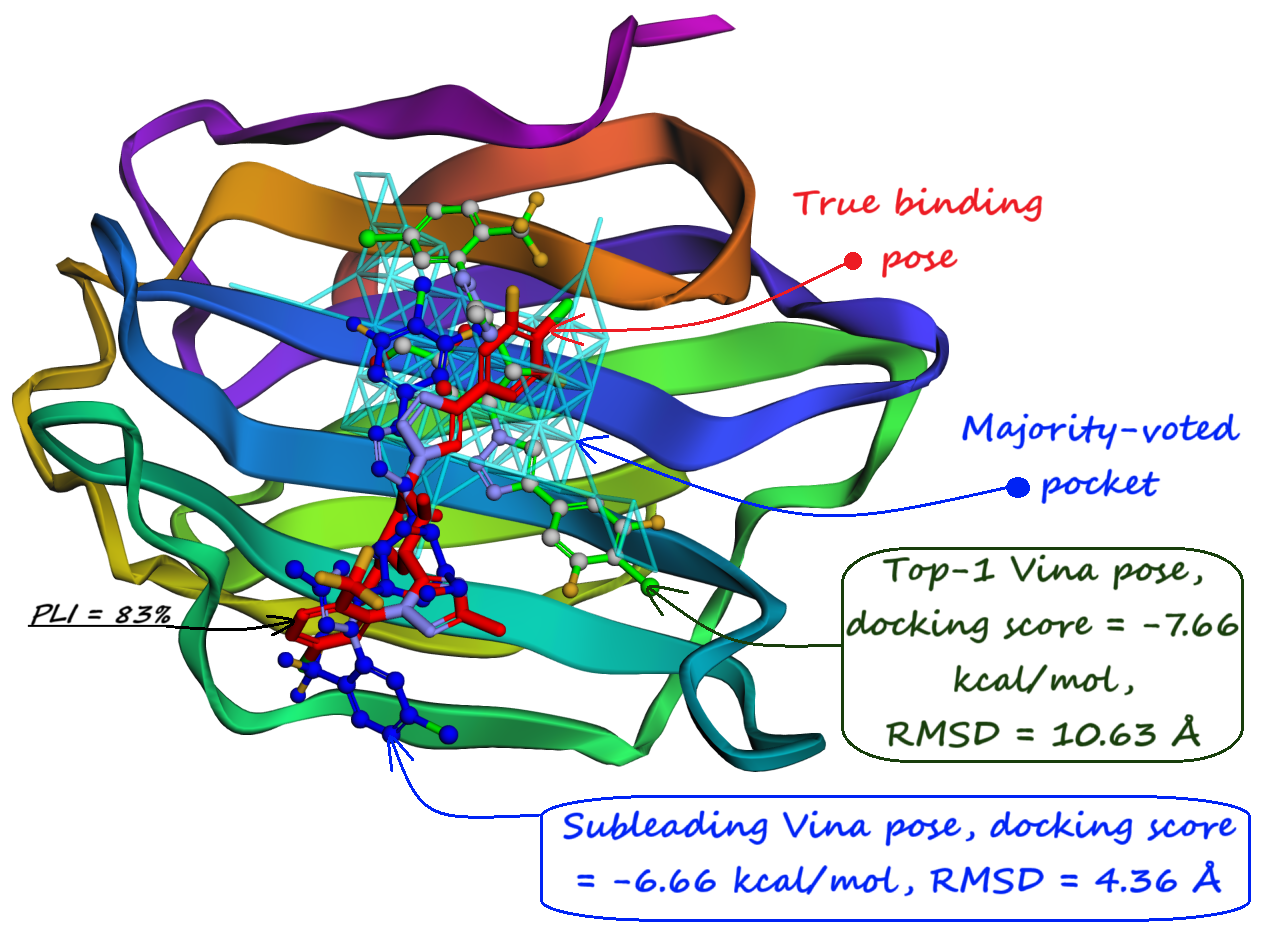}
\caption{For the 7XFA protein structure, neither the Top-1 nor any of the subleading Vina predicted poses pass the RMSD test, despite the predicted pocket covering the ligand with $\text{PLI} = 83\%$.}
\label{7XFA_D9J_Fig}
\end{figure}

Furthermore, as shown in Fig.~\ref{PB_blind_docking_results_Fig}, docking driven by individual RAPID-Net models--the ensemble components--outperforms PUResNet~\cite{kandel2021puresnet} and Kalasanty~\cite{stepniewska2020improving} in docking accuracy. Ensembling RAPID-Net models further improves docking accuracy by providing more stable results. The distribution of RMSD values for the Top-1 predicted Vina poses is shown in Fig.~\ref{PB_RMSD_Fig}. Note the issue of low recall discussed above. While the RAPID-Net ensemble predicts pockets in all 308 structures, PUResNet~\cite{kandel2021puresnet} and Kalasanty~\cite{stepniewska2020improving} predict pockets in only 278 and 263 protein structures, respectively. Since we evaluate accuracy in ``blind'' docking settings, cases with no predicted pockets are treated as failures by default.

Not all predicted pockets are positioned correctly. Of the 308 structures where RAPID-Net predicts at least one pocket, 307 contain at least one true ligand binding pose entirely within a search grid, meaning that docking can be successful in 307 out of 308 cases. In the remaining case, docking would inevitably fail due to an inaccurate search grid location. By contrast, PUResNet~\cite{kandel2021puresnet} and Kalasanty~\cite{stepniewska2020improving} have only 265 and 249 structures, respectively, where at least one true ligand pose is located completely within the search grid. This demonstrates that our model not only accurately identifies pockets in favorable protein structures but also provides reliable predictions across the entire dataset.

For comparison, Fig.~\ref{PB_130_blind_docking_results_Fig} presents the evaluation on a subset of novel protein-ligand complexes from the PoseBusters benchmark, as defined by\cite{morehead2024posebench}. Compared to the best-performing structure prediction model, AlphaFold 3, as evaluated by PoseBench, RAPID‑Net\,+\,Vina achieves a Top-1 success rate of \textbf{53.1\%}, closely approaching AF3's \textbf{59.5\%}. This result highlights RAPID-Net's ability to achieve competitive accuracy while offering a substantially lower computational cost.

\begin{figure}[]{}
\includegraphics[width=\linewidth]{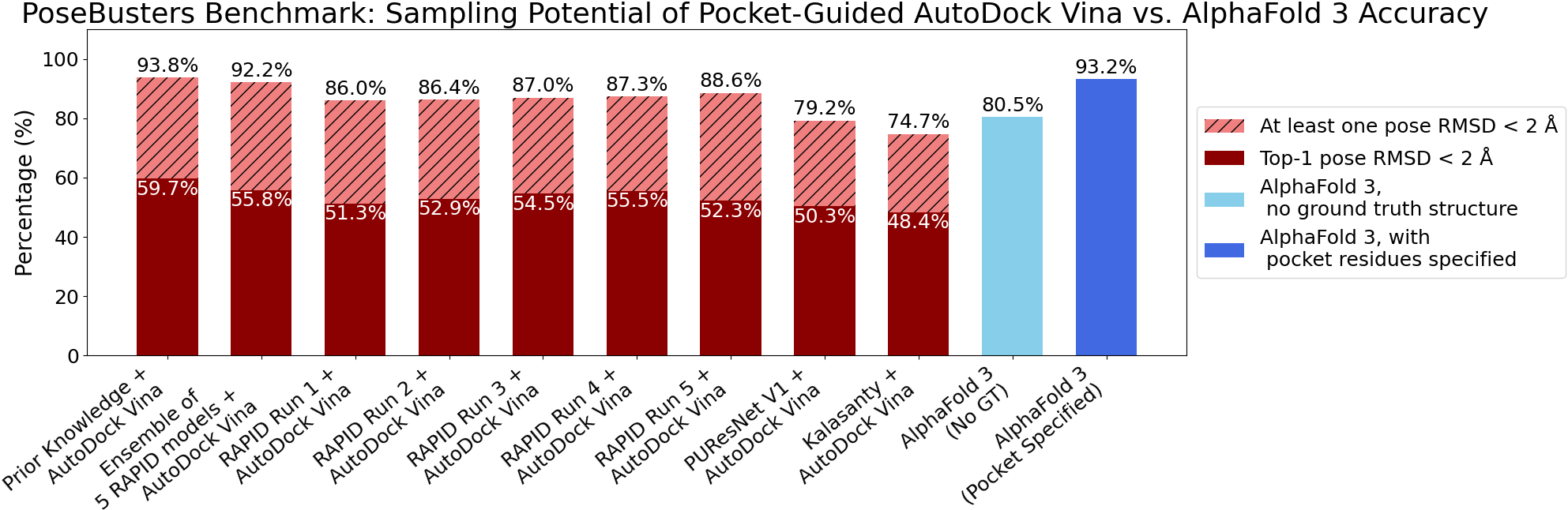}
\caption{Solid bars represent Top-1 accuracies. Dashed segments indicate the accuracy achievable when at least one pose in the ensemble is correct--potentially attainable with a more accurate reweighting tool.}
\label{PB_with_reweighting_Fig}
\end{figure}

\begin{figure*}[]{}
\includegraphics[width=\linewidth]{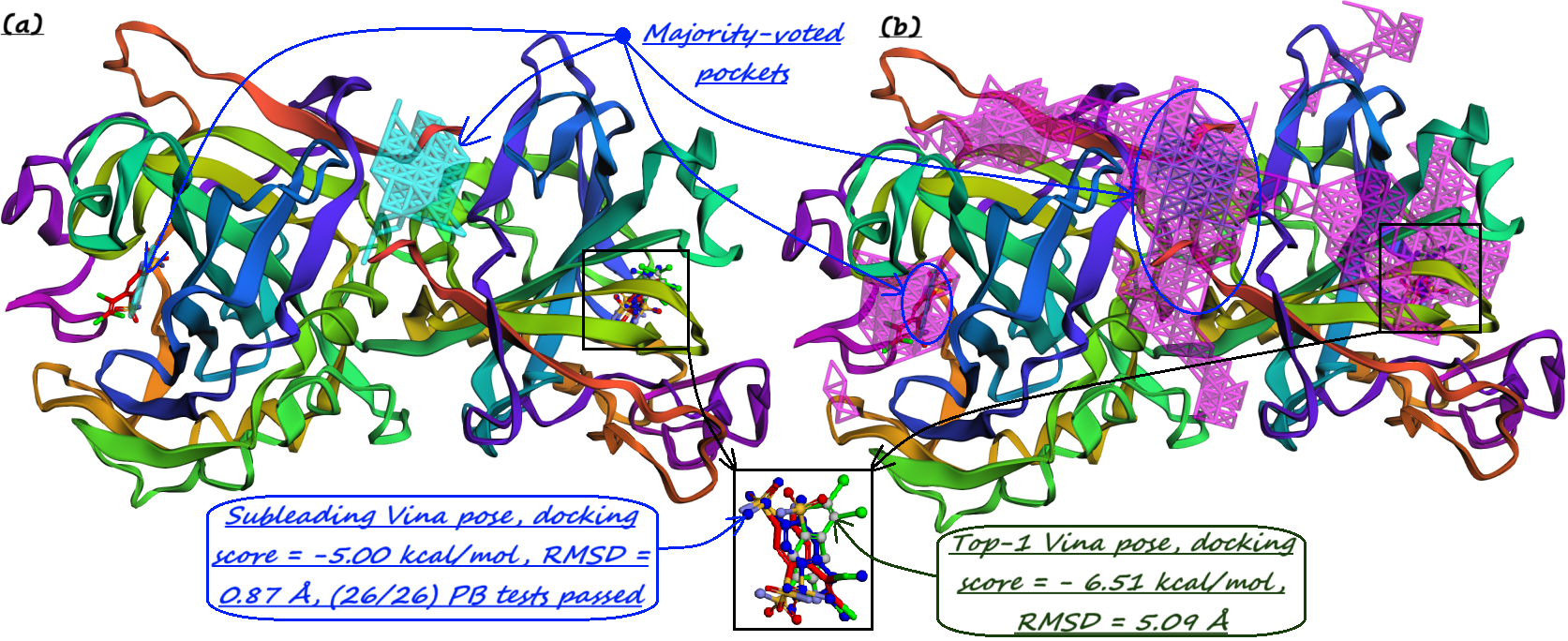}
\caption{For the 7PRI protein structure, the Top-1 Vina pose fails while a subleading one succeeds. In \textbf{(a)}, the majority-voted pockets are shown in cyan, and one of the true ligand binding poses largely overlaps with one of them. In \textbf{(b)}, minimally-reported pockets are displayed as purple sticks.}
\label{7PRI_7TI_Fig}
\end{figure*}

\begin{figure}[]{}
\includegraphics[width=\linewidth]{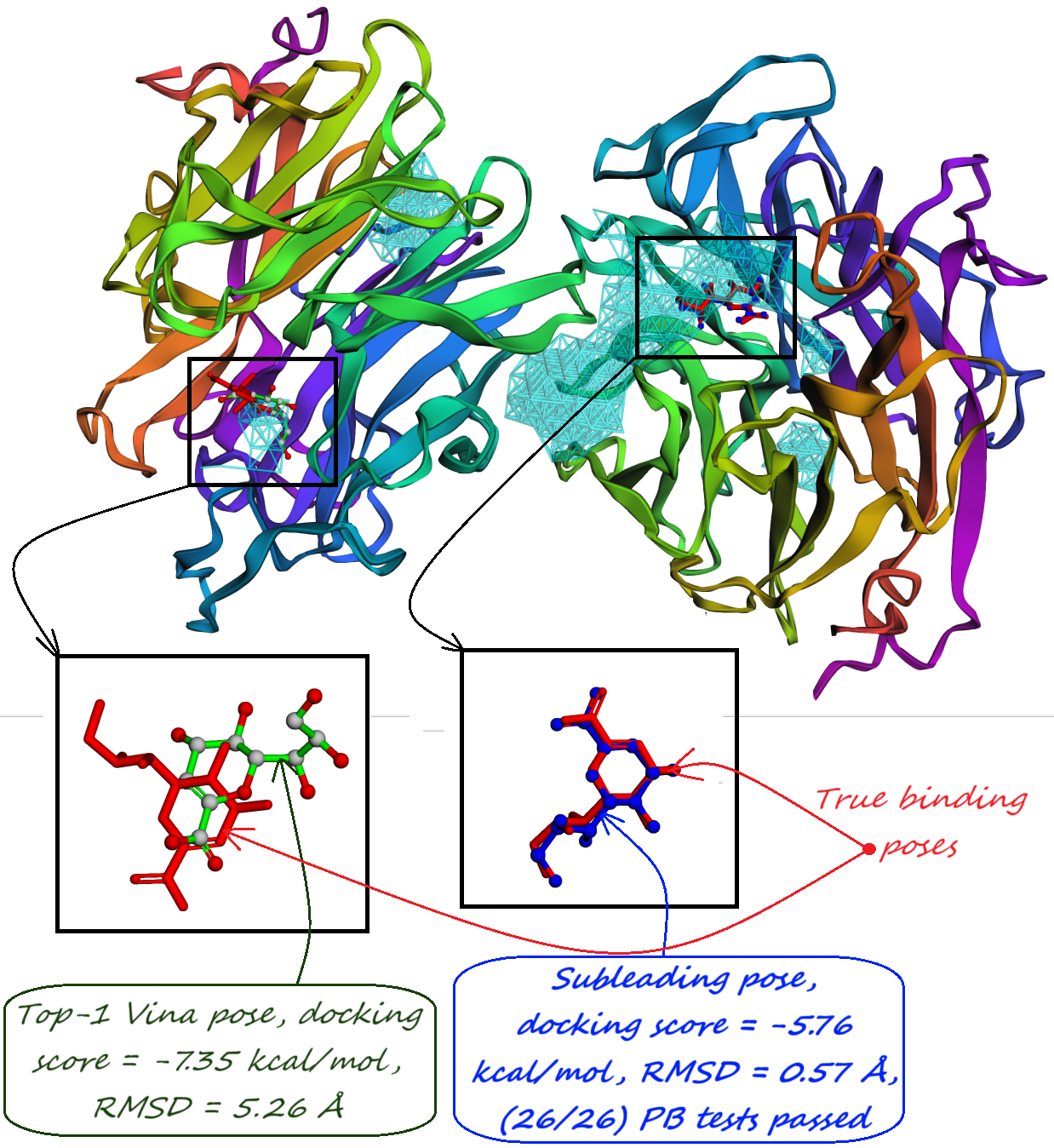}
\caption{In the 7P1F protein structure, two true ligand binding poses exist, both largely within the majority-voted pockets predicted by RAPID-Net. The Top-1 Vina pose is within one of these pockets but is inaccurate, failing the RMSD $<$ 2\,\AA\ test. However, a subleading pose occupies a different pocket and passes all tests.}
\label{7P1F_KFN_Fig}
\end{figure}
\begin{figure}[]{}
\includegraphics[width=\linewidth]{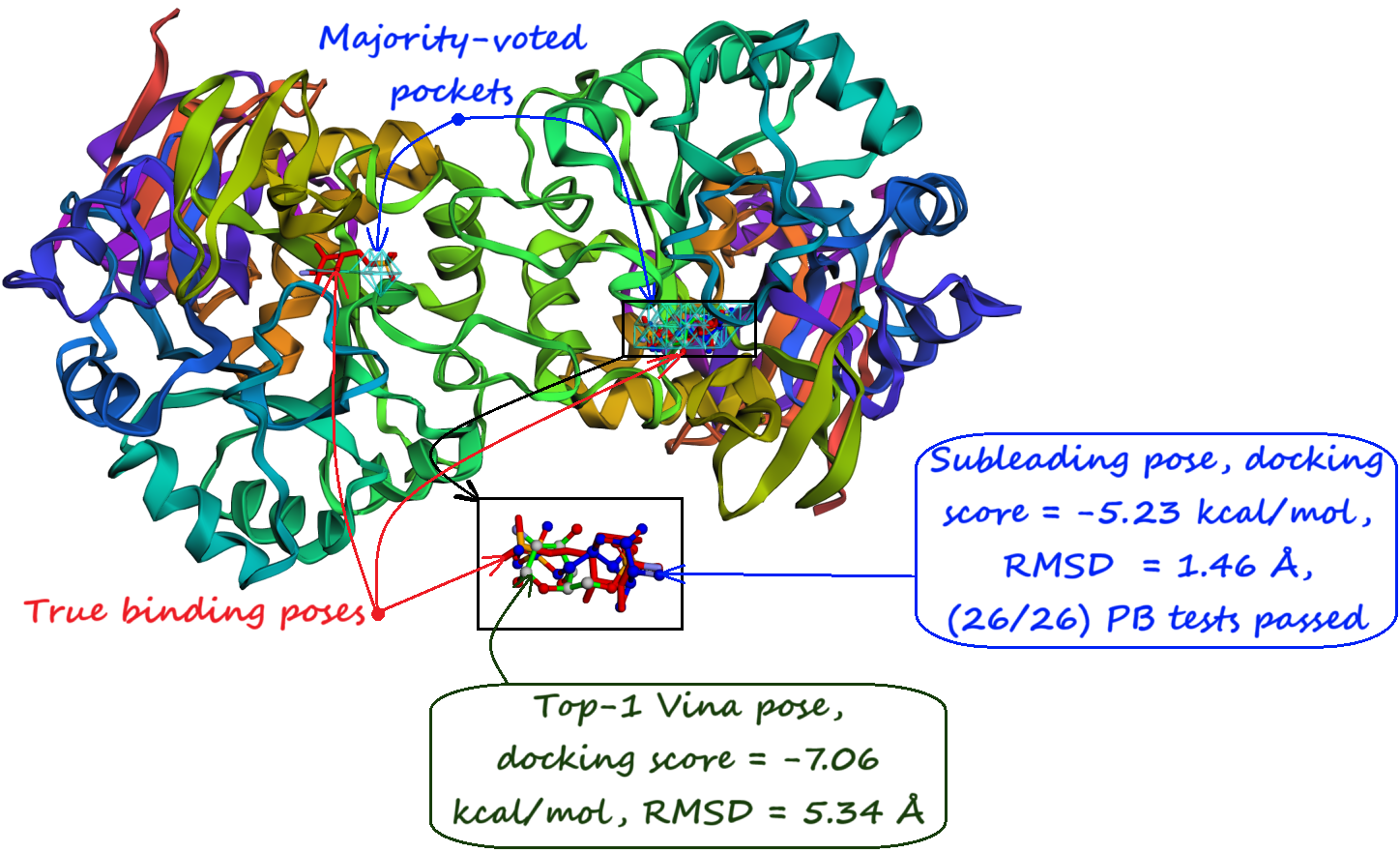}
\caption{For the 7NUT protein structure, there are two true ligand binding poses, both largely within majority-voted pockets predicted by RAPID-Net. The Top-1 Vina pose failing the RMSD $<$ 2\,\AA\ test and a subleading pose passing the test are in the same majority-voted pocket.}
\label{7NUT_GLP_Fig}
\end{figure}
\begin{figure}[]{}
\includegraphics[width=\linewidth]{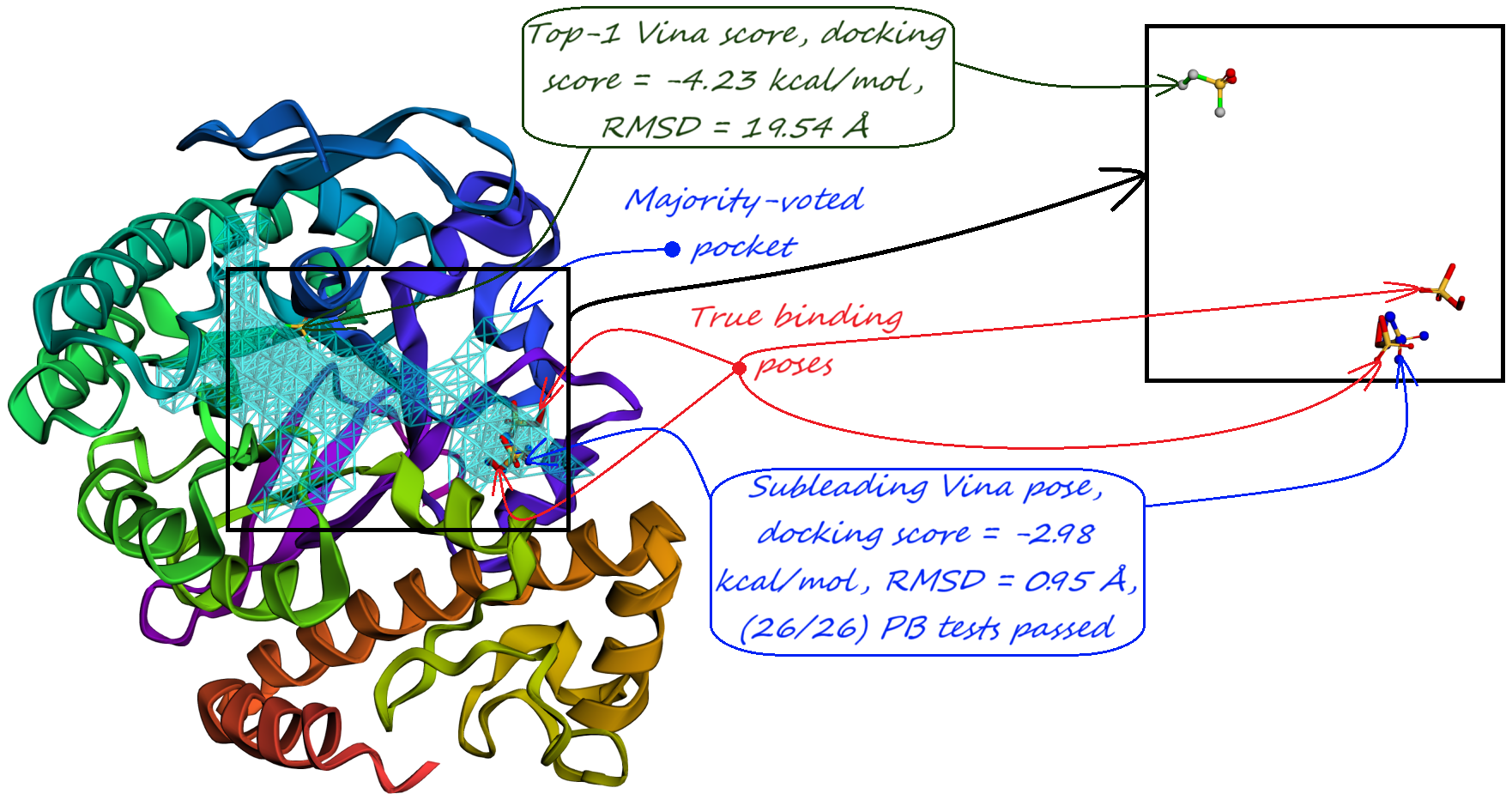}
\caption{In the 7A9E protein structure, RAPID-Net predicts a single majority-voted pocket. The Top-1 Vina pose is located in an incorrect part of that pocket, whereas a subleading pose that passes the RMSD $<$ 2\,\AA\ test--aligning with one of the true ligand binding poses--is found in another part of the same pocket.}
\label{7A9E_R4W_Fig}
\end{figure}

\begin{figure}[]{}
\includegraphics[width=\linewidth]{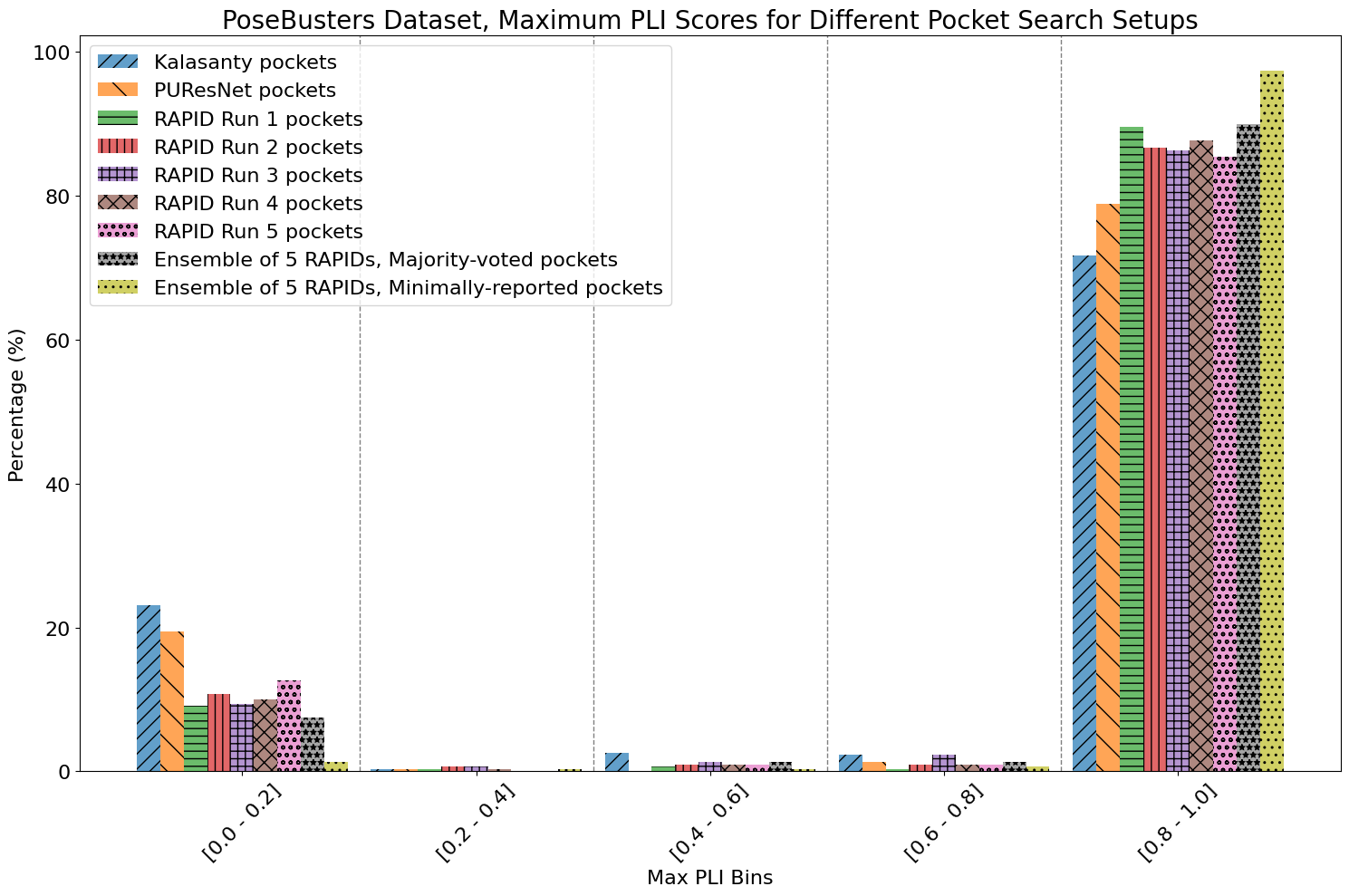}
\caption{Distribution of the maximum PLI scores corresponding to the pockets predicted by RAPID-Net, PUResNet~\cite{kandel2021puresnet}, and Kalasanty~\cite{stepniewska2020improving} for the PoseBusters dataset~\cite{Buttenschoen2024}.}
\label{PB_PLI_Fig}
\end{figure}

For the 8BTI protein structure illustrated in Fig.~\ref{8BTI_RFO_Fig}, the Top-1 Vina pose fails, but one of the subleading poses passes all tests. In contrast, for the 7XFA protein structure shown in Fig.\ref{7XFA_D9J_Fig}, neither the Top-1 nor any of the subleading Vina poses satisfy the RMSD test, with the closest pose in the ensemble having $\text{RMSD}\,=4.36\,\text{\AA}$. This occurs even though the majority-voted pocket predicted by our model covers most of the true ligand binding pose with $\text{PLI} = 83\%$.

Similarly to the 8BTI example in Fig.~\ref{8BTI_RFO_Fig}, many protein structures have the following pattern: the \textbf{Top--1} AutoDock Vina pose fails the RMSD $<$ 2\,\AA\ criterion, yet a lower-ranked pose meets it. To illustrate this behavior, we report the \emph{sampling accuracy}, defined as the fraction of cases in which \emph{at least one} pose in the ensemble achieves RMSD $< 2\,\text{\AA}$. Under this metric, as shown in Fig.~\ref{PB_with_reweighting_Fig}, AutoDock Vina with the prior knowledge of the search box~\cite{eberhardt2021autodock} has a success rate of \textbf{93.8\%}.

In comparison, AlphaFold 3, when run with pocket information available, achieves a \textbf{Top--1} accuracy of 93.2\% with RMSD~$< 2\,\text{\AA}$~\cite{abramson2024accurate}. We did not compute AlphaFold~3’s sampling accuracy and therefore make \emph{no claim} of overall superiority. This comparison merely illustrates that the current limitations in docking accuracy stem more from pose \emph{ranking} than from conformational sampling.

This difference widens in fully ``blind'' docking settings: AutoDock Vina, guided by a RAPID--Net, achieves the \textbf{92.2\%} of RMSD~$< 2\,\text{\AA}$ \emph{sampling} accuracy, whereas AlphaFold~3 achieves \textbf{80.5\%} of RMSD~$< 2\,\text{\AA}$ \textbf{Top--1} accuracy. These results suggest that improved re-ranking algorithms could further boost pocket-guided docking workflows. Furthermore, RAPID-Net performs better than PUResNet~\cite{kandel2021puresnet} and Kalasanty~\cite{stepniewska2020improving} in guiding AutoDock Vina~\cite{eberhardt2021autodock}, as can be observed in Fig.~\ref{PB_with_reweighting_Fig}.

To better illustrate this pattern, we consider four representative cases in which the \textbf{Top--1} AutoDock Vina pose fails to reach the RMSD~$< 2\,\text{\AA}$ threshold, while a pose with a much lower ranking succeeds.

Fig.~\ref{7PRI_7TI_Fig} illustrates the 7PRI protein structure, which contains two true ligand binding poses--one largely covered by the majority-voted pocket predicted by RAPID-Net, and the other by a minority-reported pocket. The Top-1 Vina pose, which fails the RMSD $<$ 2\,\AA\ test, is docked in the minority-reported pocket. A subleading pose in the same pocket passes the test, underscoring the importance of minority-reported pockets in occasionally yielding correct predictions.

Fig.~\ref{7P1F_KFN_Fig} shows the 7P1F protein structure, where both true ligand binding poses lie largely in the majority-voted pockets predicted by RAPID-Net. Although the Top-1 Vina is docked in one of these pockets and fails the RMSD $<$ 2\,\AA\ test, a subleading pose lands in another majority-voted pocket, closer to the second true ligand pose, and passes the test.

Fig.~\ref{7NUT_GLP_Fig} illustrates the 7NUT protein with two true ligand binding poses, both largely within the majority-voted pockets. Both the Top-1 Vina pose, which failed the RMSD $<$ 2\,\AA\ test, and the subleading pose, which passed the test, are located in the same majority-voted pocket near one of the true binding poses.

Finally, in the 7A9E protein structure shown in Fig.~\ref{7A9E_R4W_Fig}, RAPID-Net predicts a single majority-voted pocket. The Top-1 Vina pose is docked in the wrong part of that pocket, while a subleading pose successfully passes the RMSD $<$ 2\,\AA\ test by aligning with one of the true binding poses in another part of the pocket.

These results illustrate that the AutoDock Vina~\cite{eberhardt2021autodock} sampling mechanism is robust and reliably generates near-native binding poses within the ensemble. However, its main limitation is its ranking power: although it reliably generates poses, it struggles to distinguish correct poses from incorrect ones in its ranking. 

Finally, Fig.~\ref{PB_PLI_Fig} shows the distribution of PLI values for PUResNet~\cite{kandel2021puresnet}, Kalasanty~\cite{stepniewska2020improving}, and RAPID-Net for this dataset. All five replicas of RAPID-Net outperform both PUResNet~\cite{kandel2021puresnet} and Kalasanty~\cite{stepniewska2020improving}. Majority voting stabilizes the PLI predictions, and including minority-reported pockets further improves the PLI.

To highlight the necessity of model ensembling to provide stable pocket predictions, Table~\ref{PB_Summary_Coverage} summarizes the level of pocket and ligand overlap and lists:
\begin{itemize}

  \item The number of protein structures with at least one predicted pocket (shown in the ``Nonzero'' column). Since no protein in all our datasets is a true negative, failure to predict any pocket for a given structure is automatically an incorrect prediction.
  
  \item The number of structures with viable search grids encompassing at least one true ligand binding pose (``Within $15\,\text{\AA}$'' column), where the docking may succeed.
  
  \item The average Pocket-Ligand Intersection (PLI) value for all protein structures.
  
\end{itemize}

\begin{table}[htbp]
\centering
\small
\begin{tabularx}{\linewidth}{lccX}
\toprule
\textbf{Method}        & \textbf{Nonzero} & \textbf{Within 15 \text{\AA}} & \textbf{PLI} \\
\midrule
Kalasanty      & 263 & 249 & 74.34\% \\
PUResNet        & 278 & 265 & 79.51\% \\
RAPID Run 1     & 307 & 298 & 90.17\% \\
RAPID Run 2     & 305 & 289 & 87.90\% \\
RAPID Run 3     & 308 & 296 & 88.71\% \\
RAPID Run 4     & 308 & 300 & 88.57\% \\
RAPID Run 5     & 308 & 292 & 86.32\% \\
RAPID ensemble  & 308 & 307 & 91.44\%/98.09\% \\
\bottomrule
\end{tabularx}
\caption{Comparison of pocket predictions across different methods for PoseBusters~\cite{Buttenschoen2024} dataset consisting of 308 protein structures.}
\label{PB_Summary_Coverage}
\end{table}

For the RAPID-Net model, individual runs show variability in pocket prediction quality. For example, Run 5 predicts pockets for all 308 protein structures but with lower accuracy, resulting in the lowest average PLI of 86.32\% among the five runs. In contrast, Run 1 fails to predict pockets for one structure out of 308 but achieves the highest average PLI of 90.17\% among five runs. Combining predictions via majority voting among these RAPID-Net runs increases the average PLI to 91.44\%. Further incorporating minority-reported pockets boosts the average PLI even higher, to 98.09\%. In the final ensemble of pockets predicted by RAPID-Net, all 308 protein structures have at least one predicted pocket and only one structure lacks a search grid covering at least one true ligand binding pose. These observations highlight the importance of combining pocket predictions from multiple runs to achieve reliable results for subsequent docking.

In the next Section~\ref{Astex_Evaluation}, we perform docking on the Astex Diverse Set~\cite{hartshorn2007diverse} following a similar procedure as described for the PoseBusters~\cite{Buttenschoen2024} dataset.

\section{Evaluation on the Astex Diverse Set}
\label{Astex_Evaluation}

\begin{figure}[h]{}
\includegraphics[width=\linewidth]{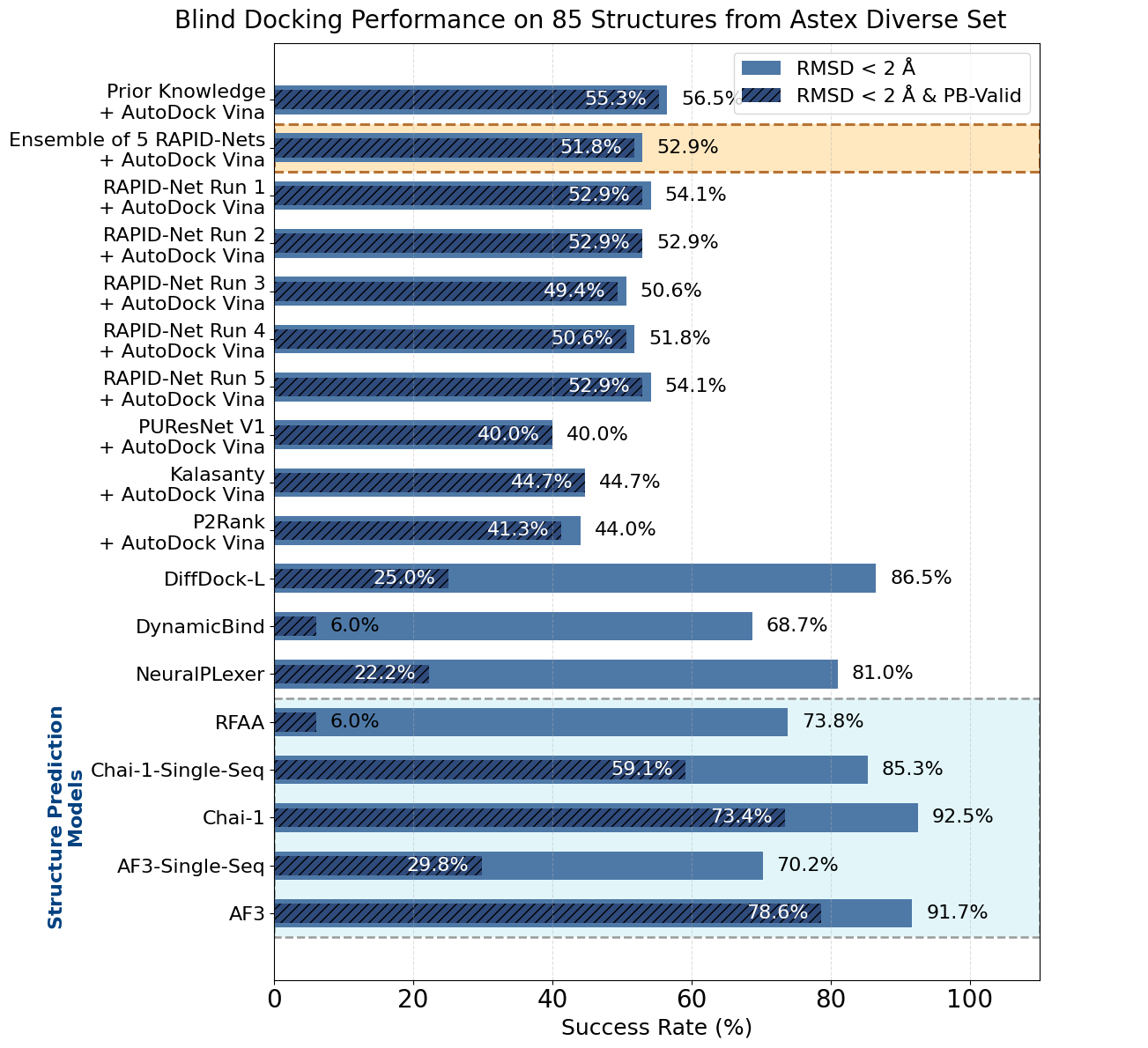}
\caption{Comparison of docking accuracies for the Astex Diverse Set~\cite{hartshorn2007diverse} when AutoDock Vina~\cite{eberhardt2021autodock} is guided by different pocket prediction algorithms.}
\label{Astex_Blind_Docking_Results_Fig}
\end{figure}

\begin{figure}[]{}
\includegraphics[width=\linewidth]{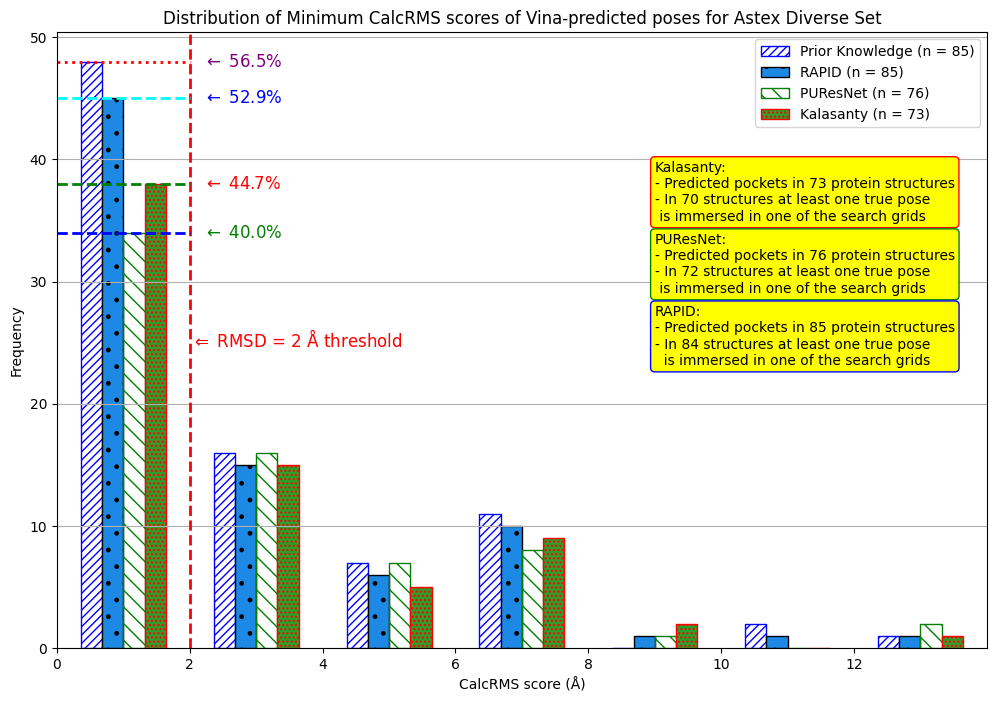}
  \caption{Distribution of RMSD values for Top-1 Vina predicted poses in the Astex Diverse Set~\cite{Buttenschoen2024}. When multiple true ligand poses are available, the RMSD to the closest one is considered.}
\label{Astex_RMSD_Fig}
\end{figure}

\begin{figure}[]{}
\includegraphics[width=\linewidth]{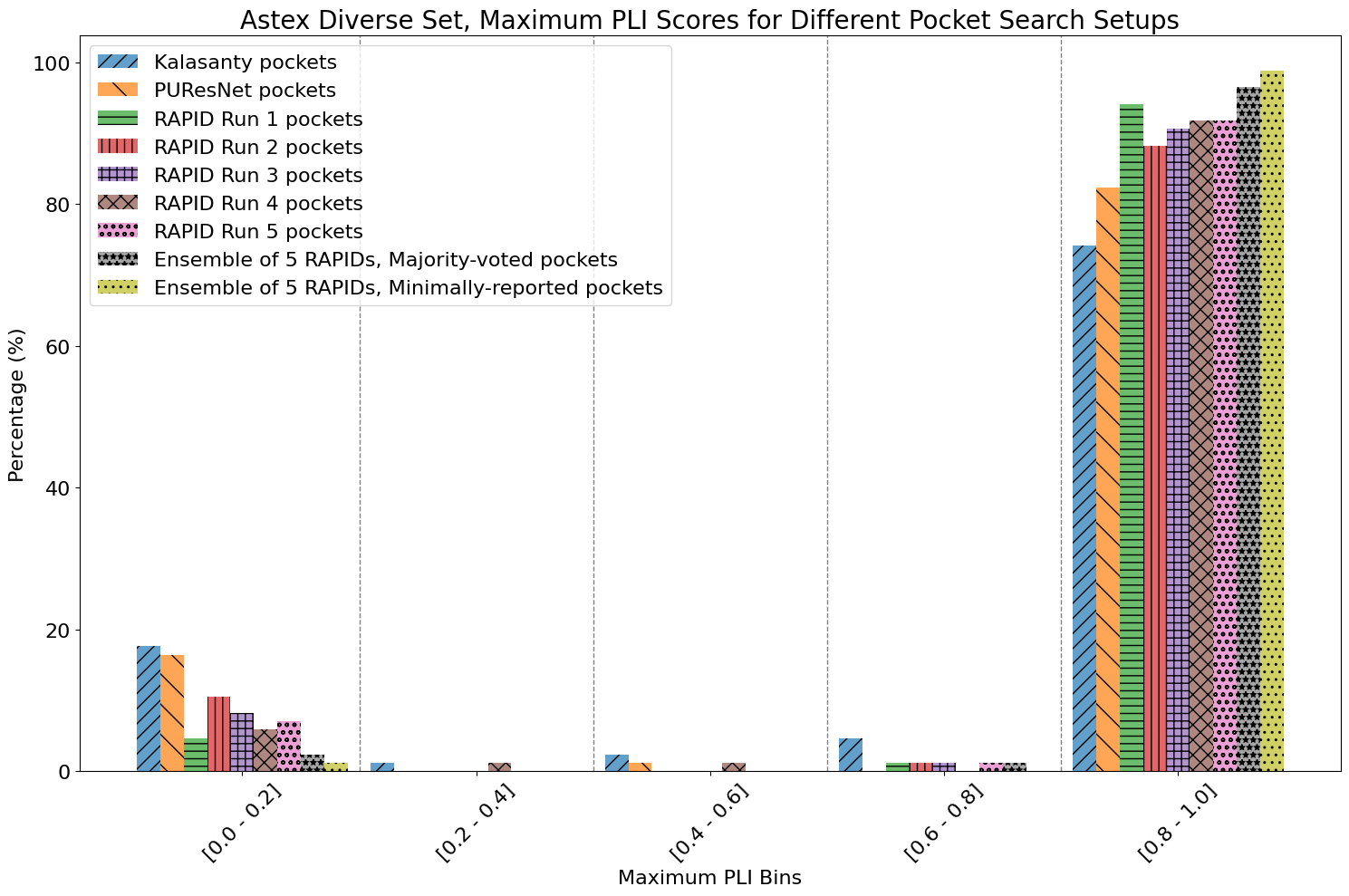}
  \caption{Distribution of the maximum PLI scores corresponding to the pockets predicted by RAPID-Net, PUResNet~\cite{kandel2021puresnet}, and Kalasanty~\cite{stepniewska2020improving} for the Astex Diverse Set~\cite{hartshorn2007diverse}.}
\label{Astex_PLI_Fig}
\end{figure}

\begin{figure}[]{}
\includegraphics[width=\linewidth]{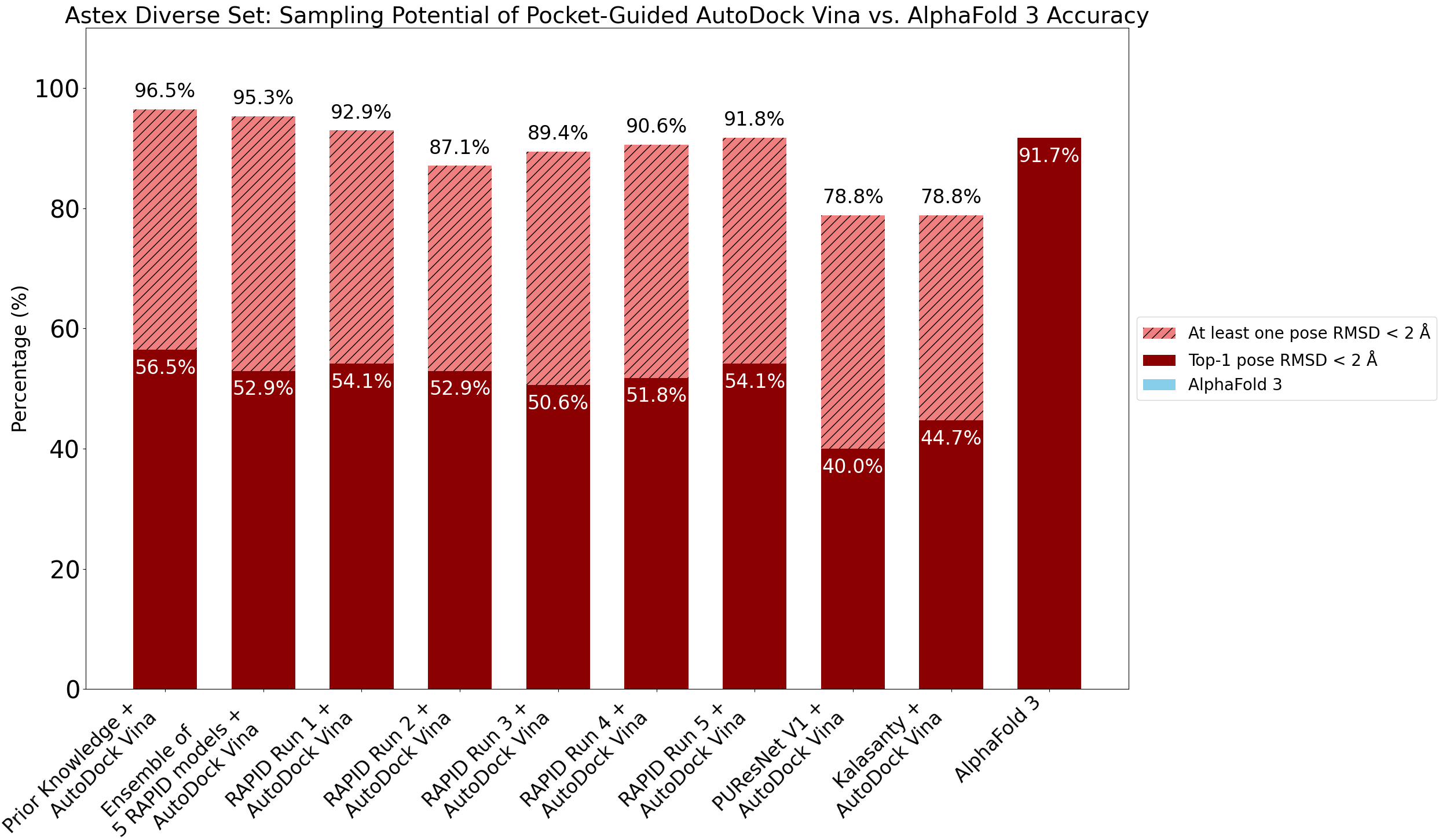}
\caption{Docking accuracy of AutoDock Vina~\cite{eberhardt2021autodock} for the Astex Diverse Set~\cite{hartshorn2007diverse} when guided by different pocket prediction algorithms, comparing Top-1 and ensemble accuracy.}
\label{Astex_Reweighted_Results_Fig}
\end{figure}

The docking results of AutoDock Vina~\cite{eberhardt2021autodock} guided by different pocket-finding algorithms for the Astex Diverse Set~\cite{hartshorn2007diverse}, consisting of 85 protein structures, are shown in Fig.~\ref{Astex_Blind_Docking_Results_Fig}. The corresponding RMSD distribution of the Top-1 predicted Vina poses is presented in Fig.~\ref{Astex_RMSD_Fig}, and the associated PLI scores are shown in Fig.~\ref{Astex_PLI_Fig}.

Several striking features emerge from these graphs. First, although each individual Run of our model outperforms the docking accuracy achieved when guiding AutoDock Vina~\cite{eberhardt2021autodock} using PUResNet~\cite{kandel2021puresnet} or Kalasanty~\cite{stepniewska2020improving}, ensembling these individual Runs does not improve the overall docking accuracy. This happens even though the ensembled version of RAPID-Net achieves a higher ligand-pocket intersection, as shown in Fig.~\ref{Astex_PLI_Fig} and Table~\ref{Astex_Summary_Coverage}. 

Furthermore, unlike the PoseBusters~\cite{Buttenschoen2024} dataset, when AutoDock Vina~\cite{eberhardt2021autodock} is guided by Kalasanty~\cite{stepniewska2020improving} on the Astex Diverse Set~\cite{hartshorn2007diverse}, it achieves better docking accuracy than when guided by PUResNet~\cite{kandel2021puresnet}, even though PUResNet~\cite{kandel2021puresnet} has more structures with at least one true ligand binding pose completely within the search grid (72 vs.\ 70, respectively).

\begin{table}[htbp]
\centering
\small
\begin{tabularx}{\linewidth}{lccX}
\toprule
\textbf{Method}        & \textbf{Nonzero} & \textbf{Within 15 \text{\AA}} & \textbf{PLI} \\
\midrule
Kalasanty      & 73 &  70 &  78.72\% \\
PUResNet       & 76 &  72 &  82.36\% \\
RAPID Run 1    & 85 &  83 &  94.83\% \\
RAPID Run 2    & 85 &  79 &  89.00\% \\
RAPID Run 3    & 85 &  78 &  91.09\% \\
RAPID Run 4    & 85 &  83 &  92.61\% \\
RAPID Run 5    & 85 &  81 &  92.47\% \\
RAPID ensemble & 85 &  84 &  97.15\%/98.82\% \\
\bottomrule
\end{tabularx}
\caption{Comparison of pocket predictions across different methods for Astex Diverse Set~\cite{hartshorn2007diverse} consisting of 85 protein structures.}
\label{Astex_Summary_Coverage}
\end{table}

\begin{figure}[]{}
\includegraphics[width=\linewidth]{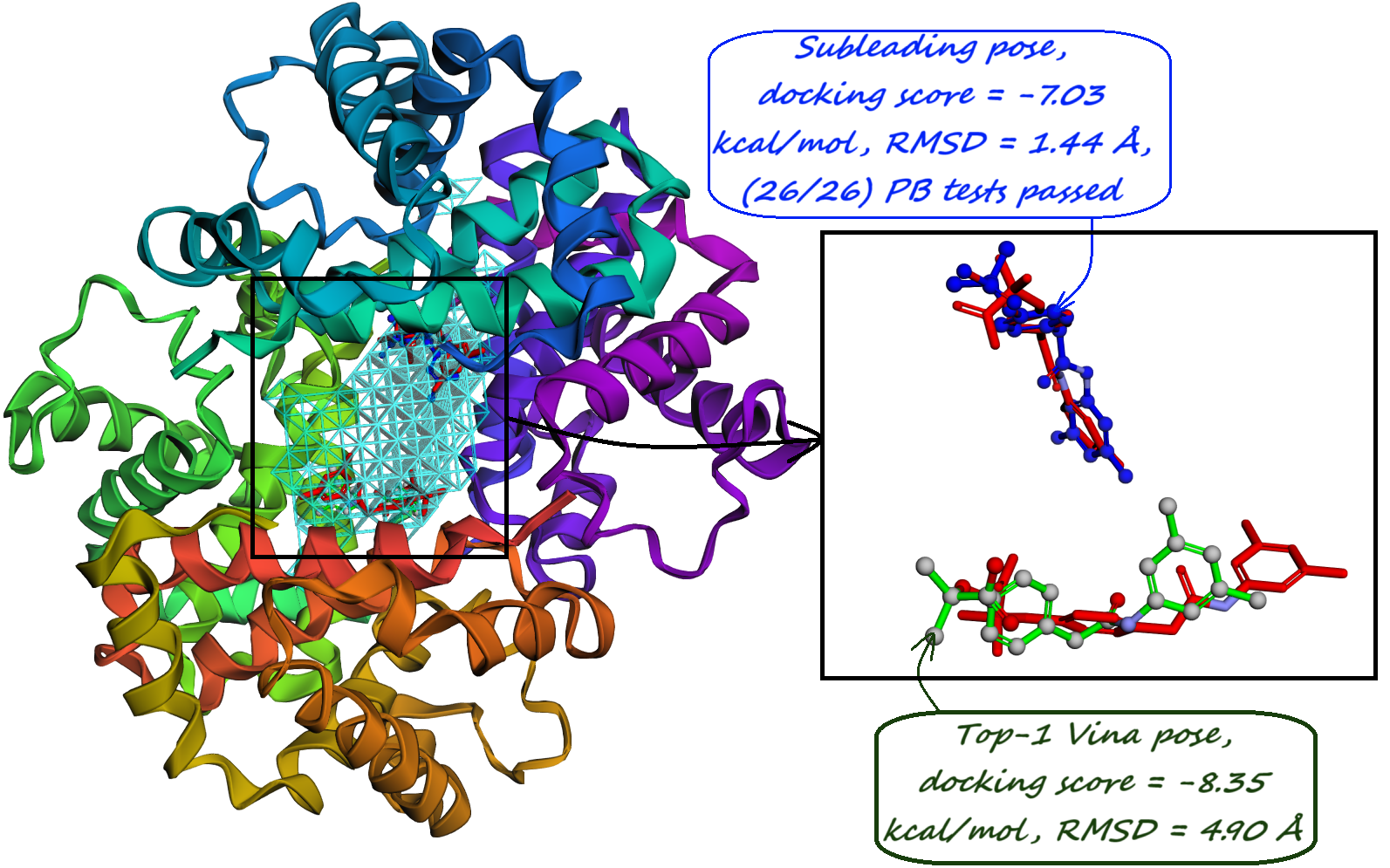}
  \caption{In the 1G9V protein structure from Astex Diverse Set~\cite{hartshorn2007diverse}, two true ligand binding poses are observed within the same majority-voted pocket predicted by RAPID-Net. While the Top-1 Vina pose fails the RMSD test, the subleading one passes all validation tests.}
\label{1G9V_RQ3_Fig}
\end{figure}

Nevertheless, when considering sampling accuracy--achieved either by the Top-1 pose or any subleading pose, as illustrated in Fig.~\ref{Astex_Reweighted_Results_Fig}--the ensembled version of RAPID-Net outperforms each of its individual Runs: 95.3\% versus 92.9\%, 87.1\%, 89.4\%, 90.6\%, and 91.8\%, respectively. Additionally, PUResNet~\cite{kandel2021puresnet} and Kalasanty~\cite{stepniewska2020improving} both achieve the same sampling accuracy of 78.8\%. A comparison of Tables~\ref{PB_Summary_Coverage} and~\ref{Astex_Summary_Coverage} alongside Figs.~\ref{PB_with_reweighting_Fig} and~\ref{Astex_Reweighted_Results_Fig} shows that RAPID-Net achieves a higher PLI rate on the Astex Diverse Set~\cite{hartshorn2007diverse} than on the PoseBusters~\cite{Buttenschoen2024} dataset. However, the Top-1 docking accuracy is higher for PoseBusters~\cite{Buttenschoen2024}, while the ``at least one correct pose in the ensemble'' accuracy is better for the Astex Diverse Set~\cite{hartshorn2007diverse}. These results further emphasize and illustrate that, in addition to accurate pocket identification, another major bottleneck in the docking process is the accurate reranking of the generated poses.

In the next Section, we compare our model against other available pocket predictors in terms of PLI metrics using their original test datasets to enable a direct side-by-side comparison.

\section{Evaluation on Coach420 and BU48 Datasets}
\label{coach420_BU48_evaluate}

For direct comparison with Kalasanty~\cite{stepniewska2020improving} and PUResNet~\cite{kandel2021puresnet}, we evaluate the corresponding PLI rates on the Coach420~\cite{Roy2012} and BU48~\cite{Huang2006} datasets. Following~\cite{kandel2021puresnet}, we exclude protein structures present in the training sets, resulting in 298 and 62 protein-ligand structures for Coach420~\cite{Roy2012} and BU48~\cite{Huang2006}, respectively.

\begin{table}[htbp]
\centering
\small
\begin{tabularx}{\linewidth}{lccX}
\toprule
\textbf{Method}        & \textbf{Nonzero} & \textbf{Within 15 \text{\AA}} & \textbf{PLI} \\
\midrule
Kalasanty      & 273 & 248  &  76.41\% \\
PUResNet       & 280 & 259  &  78.37\% \\
RAPID Run 1    & 296 & 285  &  85.68\% \\
RAPID Run 2    & 298 & 279  &  86.48\% \\
RAPID Run 3    & 293 & 264  &  76.78\% \\
RAPID Run 4    & 298 & 288  &  91.39\% \\
RAPID Run 5    & 298 & 275  &  80.59\% \\
RAPID ensemble & 298 & 297  &  86.75\%/95.49\% \\
\bottomrule
\end{tabularx}
\caption{Comparison of pocket predictions across different methods for Coach420~\cite{Roy2012} dataset consisting of 298 protein structures.}
\label{Coach420_Summary_Coverage}
\end{table}

\begin{table}[htbp]
\centering
\small
\begin{tabularx}{\linewidth}{lccX}
\toprule
\textbf{Method}        & \textbf{Nonzero} & \textbf{Within 15 \text{\AA}} & \textbf{PLI} \\
\midrule
Kalasanty      & 54 & 53  &  77.80\% \\
PUResNet       & 42 & 39  &  46.90\% \\
RAPID Run 1    & 61 & 60  &  86.52\% \\
RAPID Run 2    & 62 & 62  &  97.57\% \\
RAPID Run 3    & 61 & 60  &  86.68\% \\
RAPID Run 4    & 60 & 58  &  84.47\% \\
RAPID Run 5    & 62 & 43  &  46.97\% \\
RAPID ensemble & 62 & 62  &  95.79\%/99.82\% \\
\bottomrule
\end{tabularx}
\caption{Comparison of pocket predictions across different methods for BU48~\cite{Huang2006} dataset consisting of 62 protein structures.}
\label{BU48_Summary_Coverage}
\end{table}

\begin{figure}[]{}
\includegraphics[width=\linewidth]{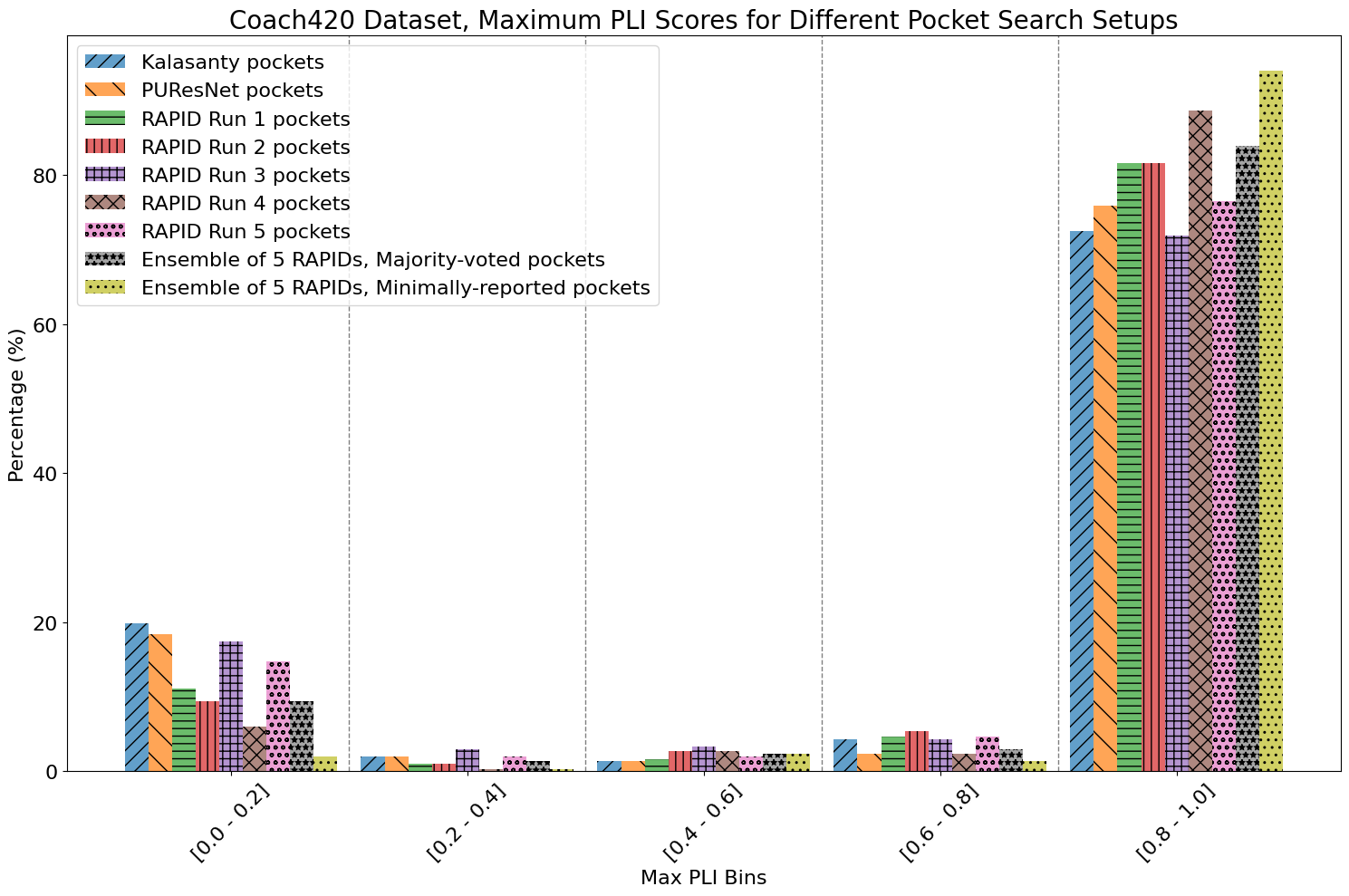}
  \caption{Distribution of the maximum PLI scores corresponding to the pockets predicted by RAPID-Net, PUResNet~\cite{kandel2021puresnet}, and Kalasanty~\cite{stepniewska2020improving} for the Coach420~\cite{Roy2012} dataset.}
\label{Coach420_Fig}
\end{figure}

\begin{figure}[]{}
\includegraphics[width=\linewidth]{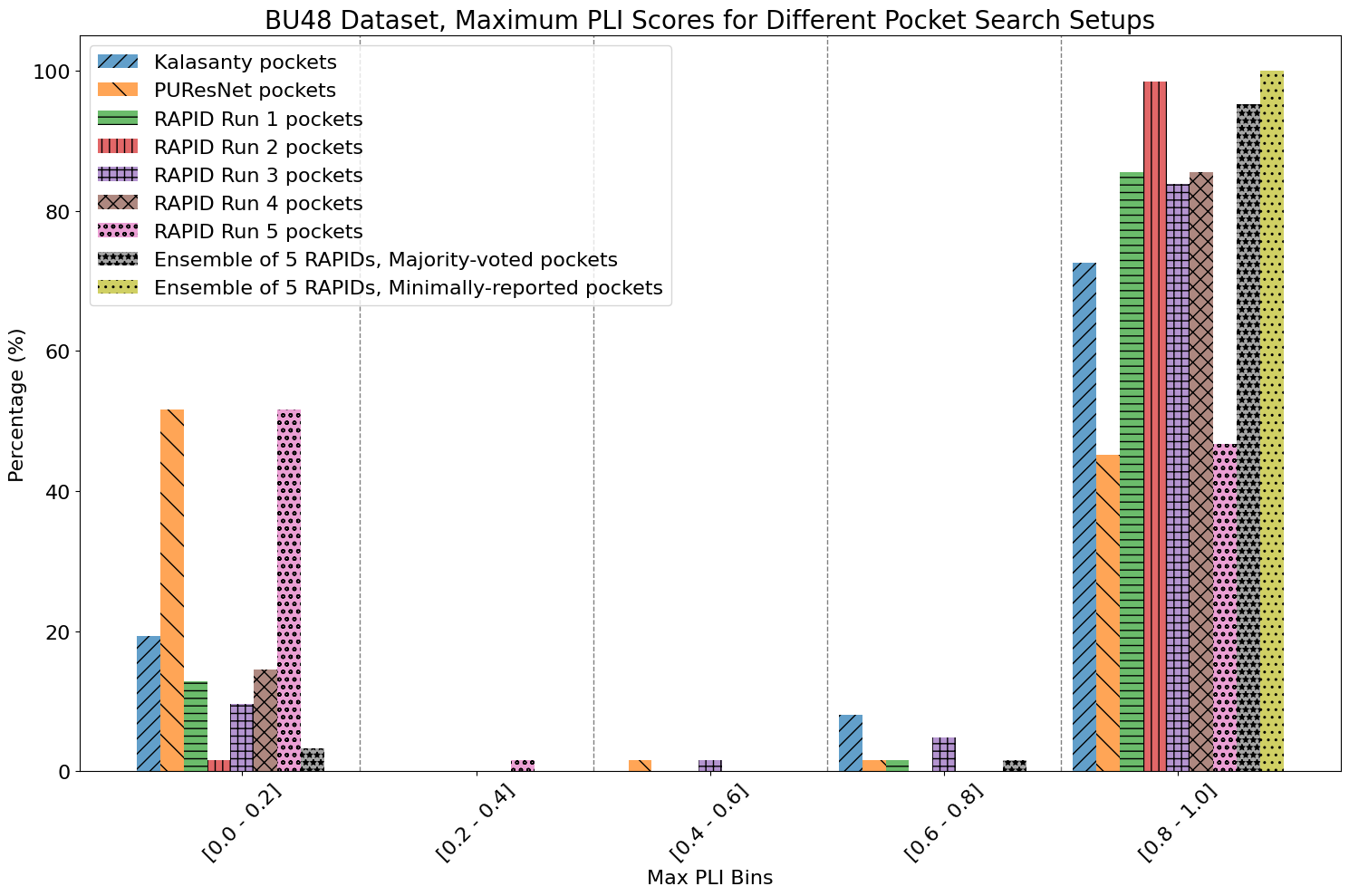}
  \caption{Distribution of the maximum PLI scores corresponding to the pockets predicted by RAPID-Net, PUResNet~\cite{kandel2021puresnet}, and Kalasanty~\cite{stepniewska2020improving} for the BU48~\cite{Huang2006} dataset.}
\label{BU48_Fig}
\end{figure}

\begin{figure*}[]{}
  \centering
  \includegraphics[width=\linewidth]{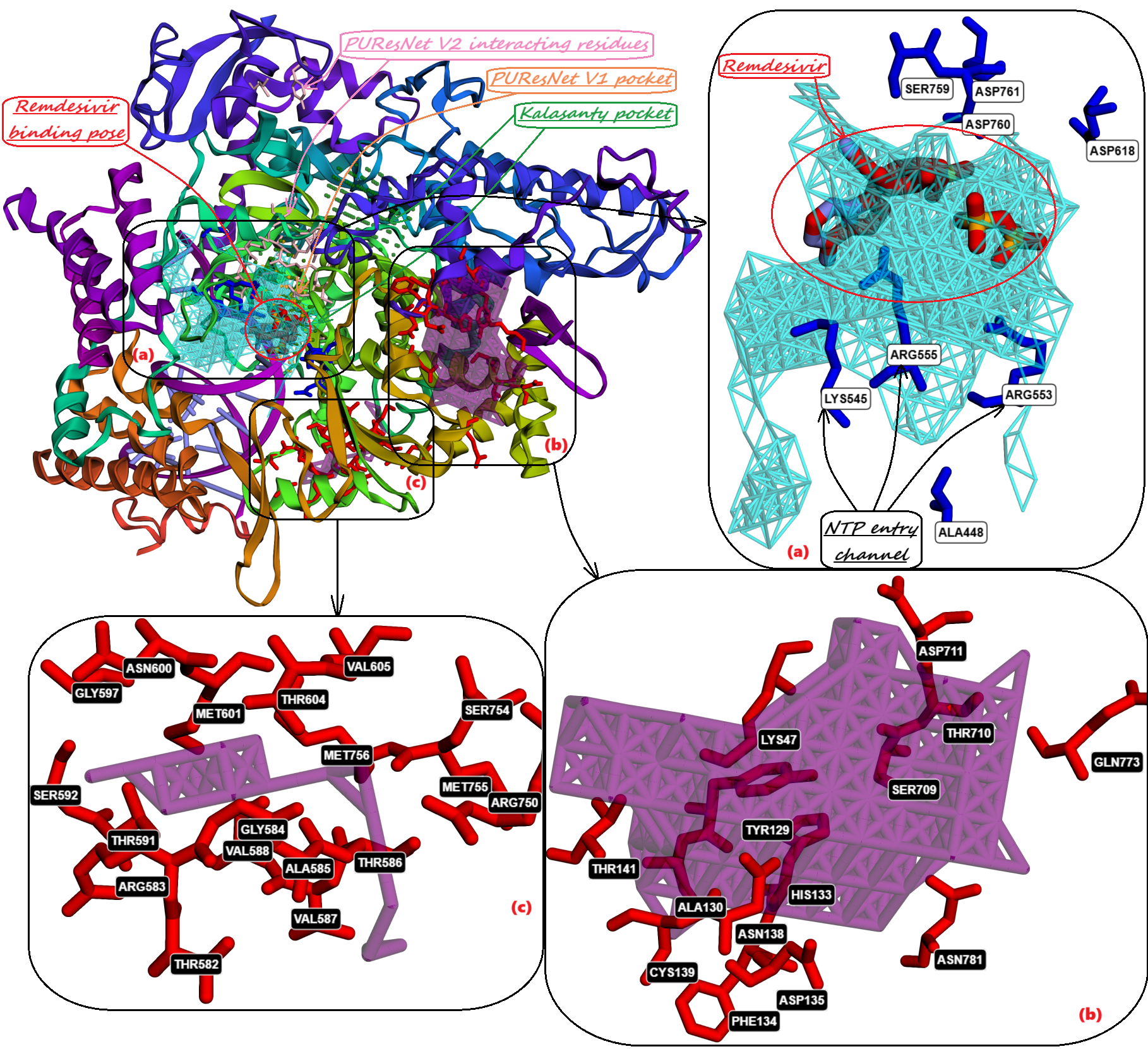}

  \caption{For the \textbf{Nsp12} protein~\cite{yin2020structural}, the pockets predicted by Kalasanty~\cite{stepniewska2020improving} and PUResNet~\cite{kandel2021puresnet} -- shown as green and orange dots, respectively -- and the likely interacting residues predicted by PUResNet~V2~\cite{jeevan2024puresnetv2} correspond to the orthosteric remdesivir binding site~\cite{yin2020structural}. In contrast, RAPID-Net not only identifies the orthosteric site -- corresponding to the majority-voted pocket, shown as cyan sticks -- but also successfully detects putative secondary and allosteric pockets reported in previous studies~\cite{faisal2021computational,parvez2020prediction}, which correspond to the minimally-reported pockets predicted by our model, shown as purple sticks. }

  \label{Nsp12_Fig}
\end{figure*}

\begin{figure}[h]{}
\includegraphics[width=\linewidth]{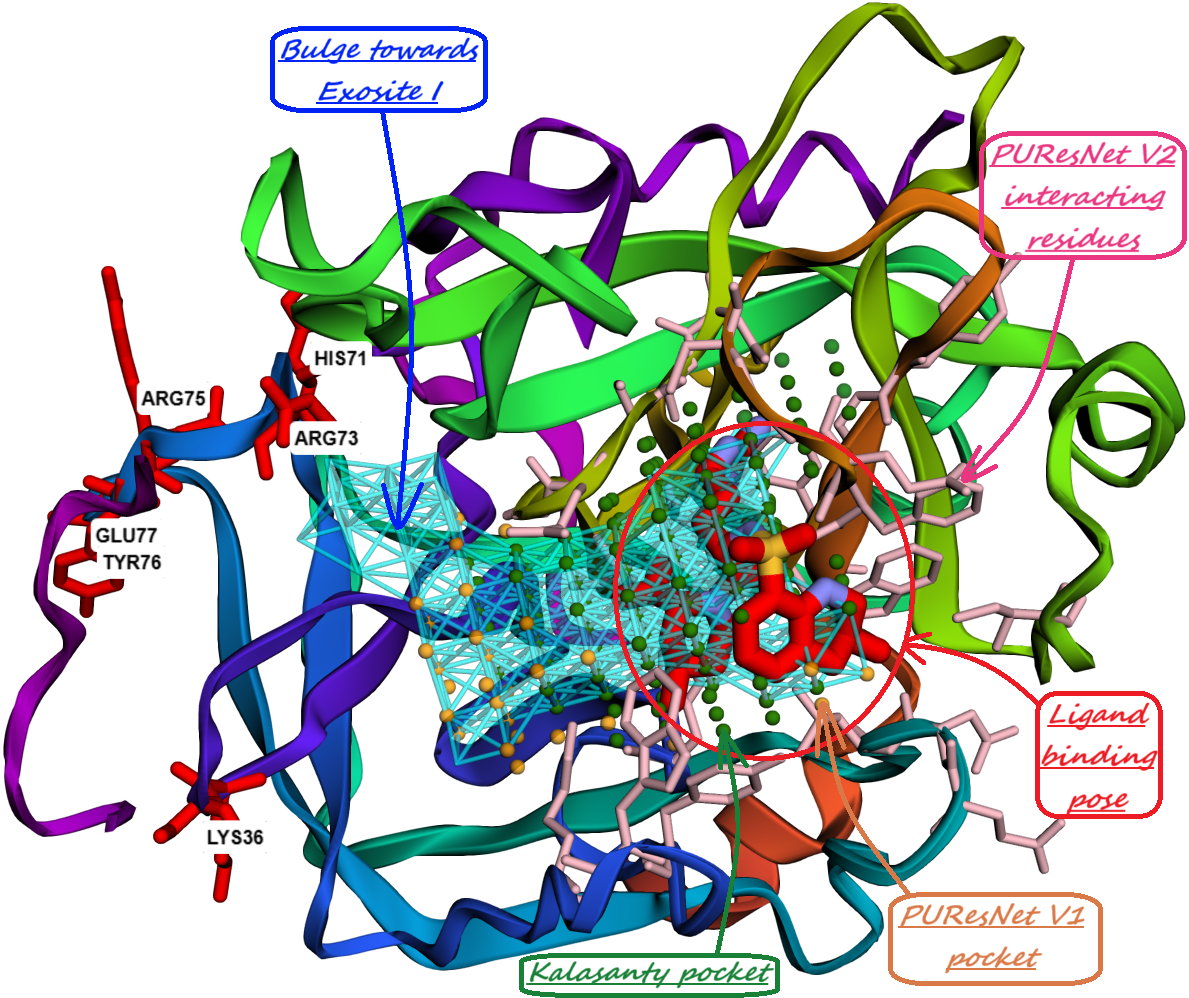}
  \caption{\textbf{Thrombin (RCSB PDB: 1DWC).} Unlike the pockets predicted by Kalasanty~\cite{stepniewska2020improving} and PUResNet~\cite{kandel2021puresnet} --shown as green and orange dots, respectively -- the majority-voted pocket predicted by our model, shown as cyan sticks, has a distinct bulge extending toward residues associated with Exosite~\rm{I}.}
  \label{1DWC_Fig}
\end{figure}

\begin{figure*}[]
  \centering
  \begin{subfigure}[b]{0.48\textwidth}
    \includegraphics[width=\linewidth]{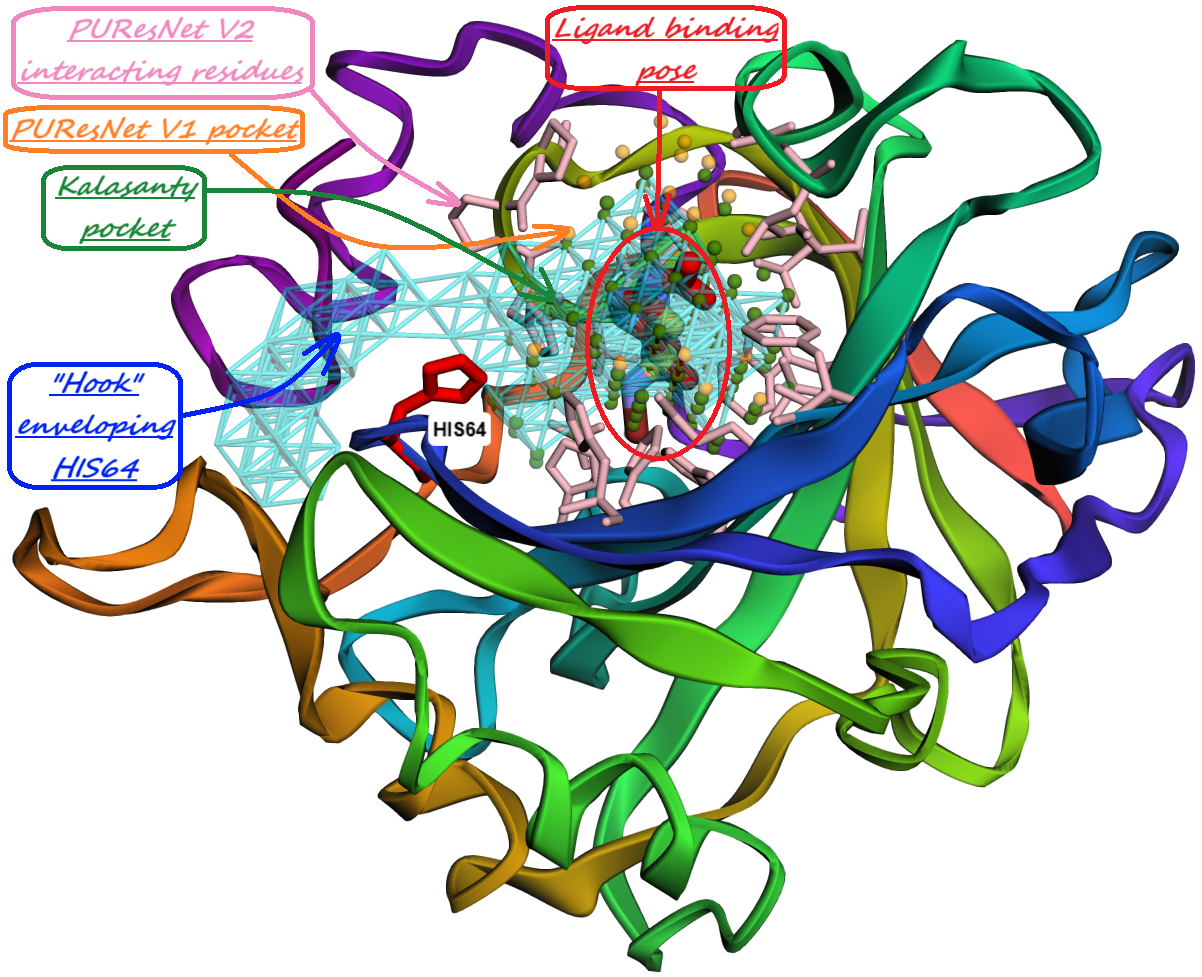}
    \caption{\textbf{Human Carbonic Anhydrase I (RCSB PDB: 1AZM).} The majority-voted pocket predicted by our model reveals a distinct hook-like bulge that wraps around His64 residue, responsible for the protein transfer to the main binding site.}
    \label{1AZM_Fig}
  \end{subfigure}
  \hfill
  \begin{subfigure}[b]{0.48\textwidth}
    \includegraphics[width=\linewidth]{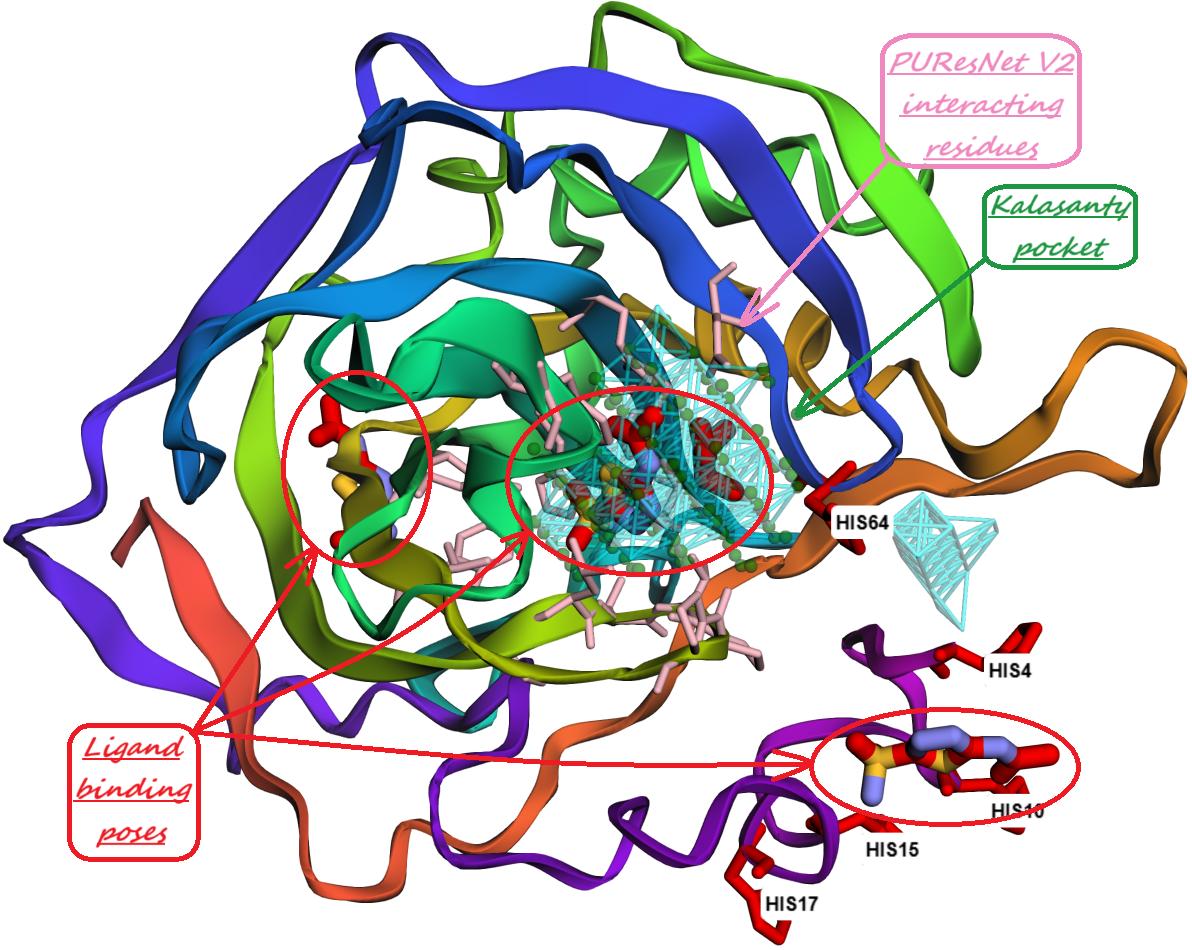}
    \caption{\textbf{Human Carbonic Anhydrase II (RCSB PDB: 3HS4).} PUResNet~\cite{kandel2021puresnet} does not predict any pockets for this protein structure, but Kalasanty~\cite{stepniewska2020improving} predicts one around the main binding site. One majority-voted pocket predicted by our model covers the active site and the other one is located near the histidine cluster responsible for the proton shuttle.}
    \label{3HS4_1_Fig}
  \end{subfigure}
    \begin{subfigure}[b]{0.48\textwidth}
    \includegraphics[width=\linewidth]{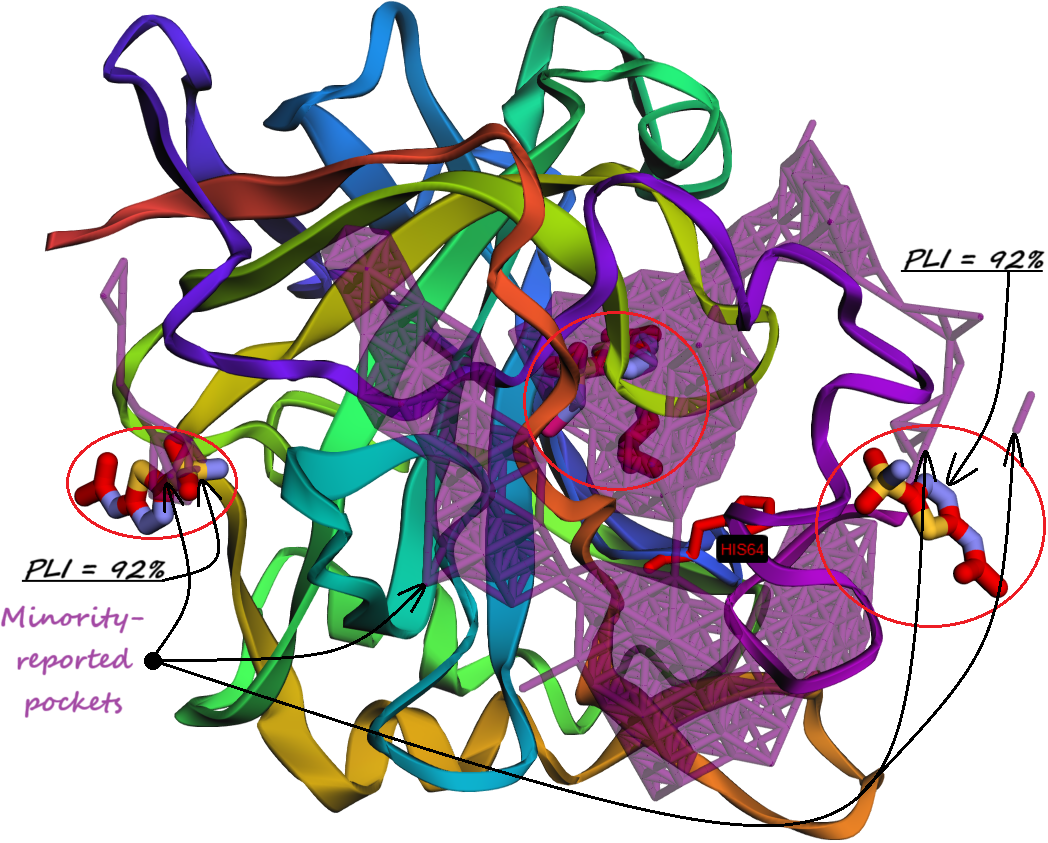}
    \caption{\textbf{Human Carbonic Anhydrase II (RCSB PDB: 3HS4).} Minority-reported pockets predicted by our model not only cover the active site but also have a $\text{PLI} = 92\%$ with the two AZM binding poses in shallow pockets on the enzyme's surface.}
    \label{3HS4_3_Fig}
  \end{subfigure}
  \hfill
  \begin{subfigure}[b]{0.48\textwidth}
    \includegraphics[width=\linewidth]{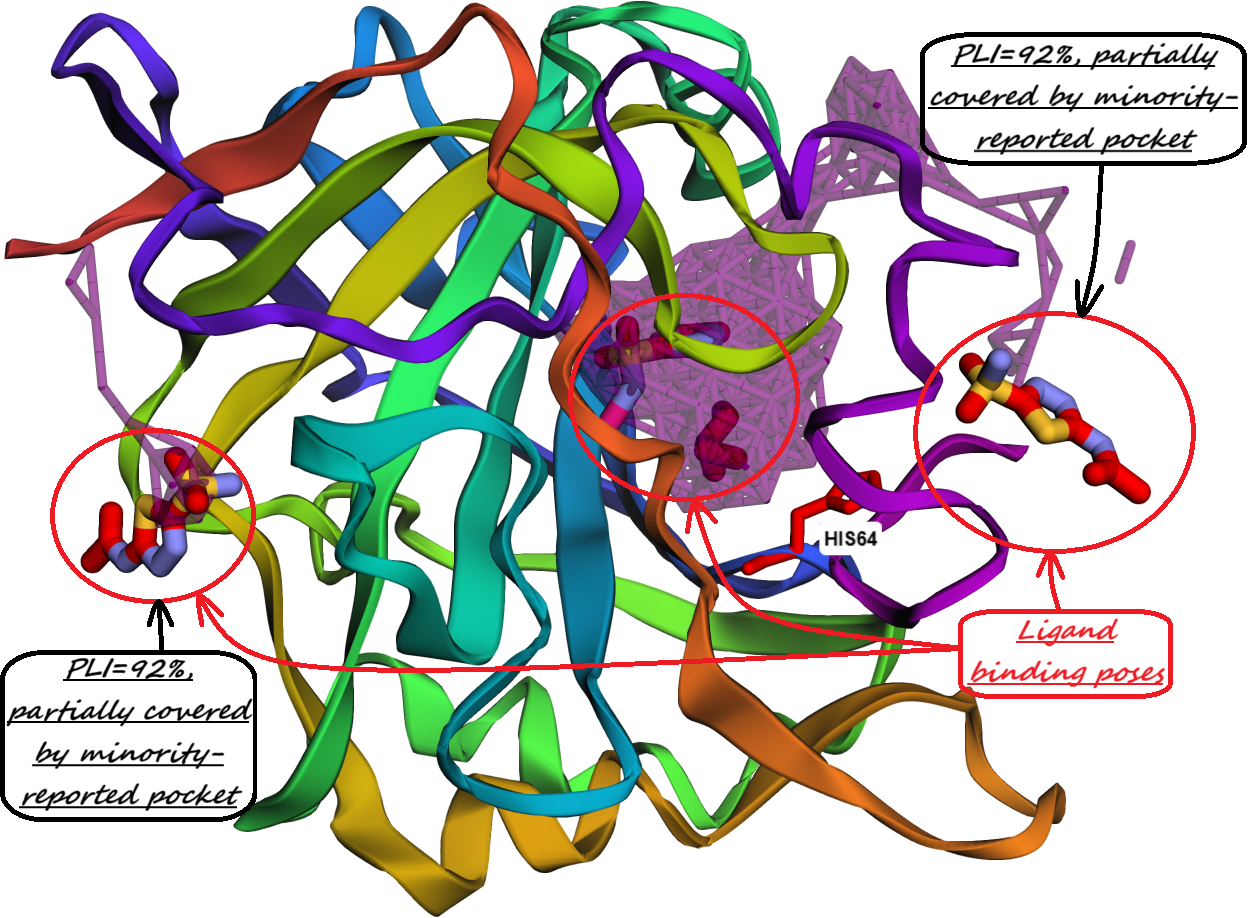}
    \caption{\textbf{Human Carbonic Anhydrase II (RCSB PDB: 3HS4).} The interacting residues predicted by PUResNet V2~\cite{jeevan2024puresnetv2} are located only near the active site, with secondary binding sites not being predicted.}
    \label{3HS4_2_Fig}
  \end{subfigure}
  \caption{Majority-voted pockets predicted by RAPID-Net are represented by cyan sticks, while the minimally-reported pockets are shown by purple sticks. The interacting residues predicted by PUResNet V2~\cite{jeevan2024puresnetv2} are shown in yellow.}
  \label{combined_fig}
\end{figure*}

\begin{figure*}[]{}
  \centering
  \includegraphics[width=\linewidth]{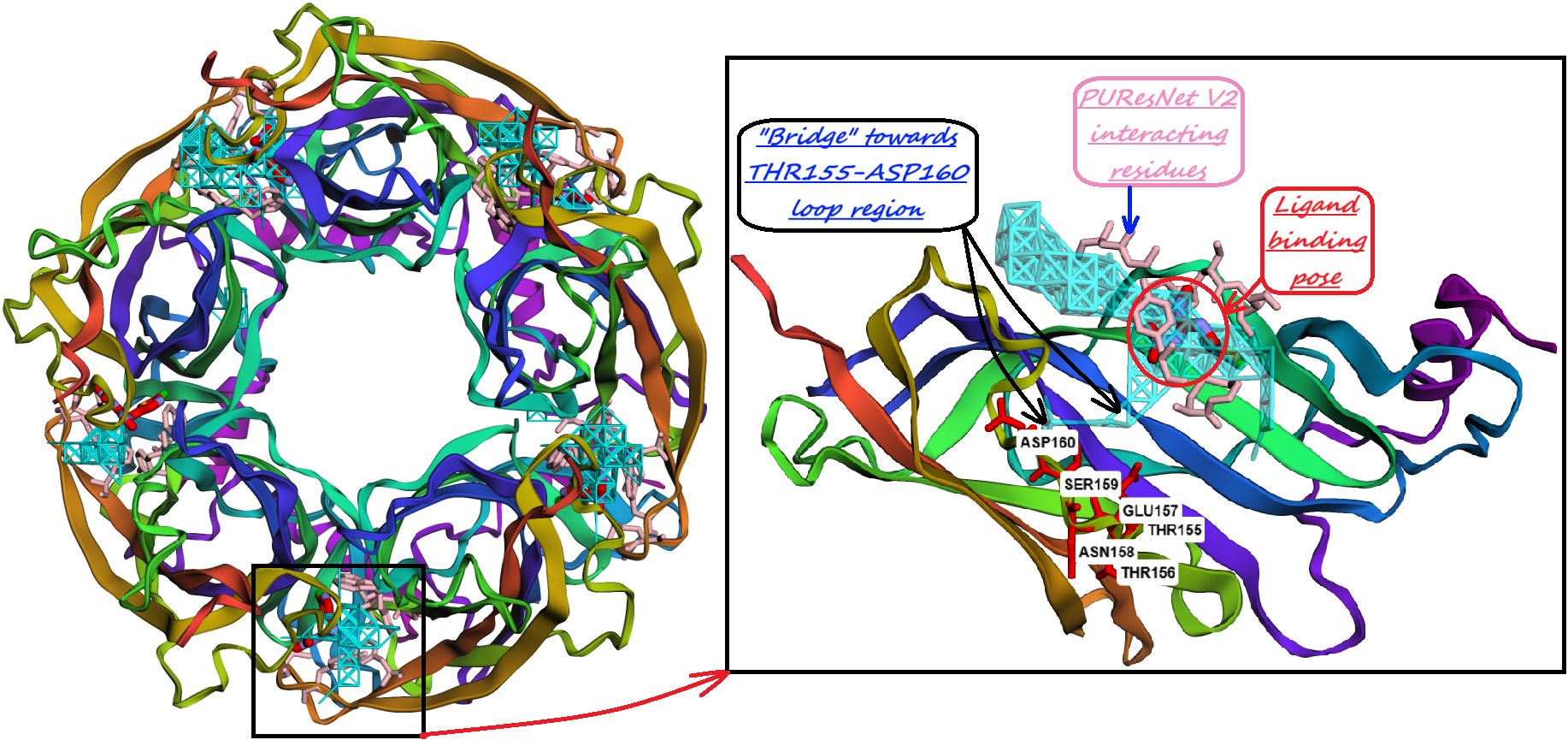}
  \caption{\textbf{Ls-AChBP (RCSB PDB: 2ZJV).} The majority-voted pocket predicted by our model has a distinct bridge towards the Thr155-Asp160 region belonging to the F loop. Kalasanty~\cite{stepniewska2020improving} and PUResNet V1~\cite{kandel2021puresnet} do not predict any pockets for this protein structure, while PUResNet V2~\cite{jeevan2024puresnetv2} predicts interacting residues only in the immediate vicinity of the binding pose.}
  \label{2ZJV_1_Fig}
\end{figure*}

Unlike the original PUResNet~\cite{kandel2021puresnet} paper where PLI values were evaluated only on protein structures where the distance-center-to-center (DCC) between predicted pocket centers and ligand centers was $\le 4\text{\AA}$, we report the results for all protein structures, over the whole dataset. This comparison seems more appropriate because a rough initial approximation to set up the search grid is often sufficient to achieve successful docking, as illustrated in Fig.~\ref{8FAV_4Y5_Fig}. Furthermore, as discussed in the next Section, the RAPID-Net model sometimes predicts meaningful ``tunnels'' or ``bridges'' to distant binding sites that indirectly influence ligand binding. In such cases, the DCC metric becomes irrelevant since it does not reflect the functional importance of these remote interactions.

Table~\ref{Coach420_Summary_Coverage} and Fig.~\ref{Coach420_Fig} summarize the results for the Coach420 dataset~\cite{Roy2012}. Except for Run 3, every RAPID-Net run surpasses PUResNet~\cite{kandel2021puresnet} and Kalasanty~\cite{stepniewska2020improving} in terms of an average PLI score. Furthermore, when comparing the number of protein structures in which at least one true ligand binding pose is entirely within the search grid, up to the largest grid with a threshold of $15\ \text{\AA}$, all RAPID-Net Runs outperform both PUResNet~\cite{kandel2021puresnet} and Kalasanty~\cite{stepniewska2020improving}. Similarly, Table~\ref{BU48_Summary_Coverage} and Fig.~\ref{BU48_Fig} present the results for the BU48 dataset~\cite{Huang2006}. Here, all RAPID-Net Runs, except Run 5, outperform both Kalasanty and PUResNet in terms of an average PLI and the number of protein structures where at least one true ligand binding pose is completely contained within the search grids.

Different RAPID-Net Runs have varying performances across the four test datasets, reflecting the inherent element of randomness in the model training. For example, on the PoseBusters~\cite{Buttenschoen2024} dataset, Run 2 has the fewest protein structures with viable predicted search grids, as shown in Table~\ref{PB_Summary_Coverage}. In contrast, Run 3 produces the fewest viable grids for the Astex Diverse Set~\cite{hartshorn2007diverse} and Coach420~\cite{Roy2012}, as shown in Tables~\ref{Astex_Summary_Coverage} and~\ref{Coach420_Summary_Coverage}, while Run 5 shows the lowest PLI for BU48~\cite{Huang2006}, as shown in Table~\ref{BU48_Summary_Coverage}. Notably, despite having lower coverage on Coach420~\cite{Roy2012} and BU48~\cite{Huang2006}, Run 5 achieves the highest ``at least one correct pose in the ensemble'' rate on PoseBusters~\cite{Buttenschoen2024}, as shown in Fig.~\ref{PB_with_reweighting_Fig}.

These results further emphasize the importance of ensembling model predictions to capture all potential binding sites and account for variability in model training. By combining five RAPID-Net models, we mitigate performance defects of individual model Runs, yielding more robust and reliable results. At the same time, as can be observed from all the data that we presented, across all four test datasets, RAPID-Net consistently exhibits stronger generalization ability than both PUResNet~\cite{kandel2021puresnet} and Kalasanty~\cite{stepniewska2020improving}.

In addition to docking accuracy and the coverage of true ligand binding poses by predicted pockets, it is crucial to identify relevant distant sites that indirectly influence ligand binding, which is the topic of the next Section.

\section{Identification of Allosteric Sites, Exosites, and Flexible Regions for Drug Design}
\label{Distant_Sites}

As noted earlier, a common problem with Deep Learning-based pocket prediction methods is that they may miss secondary pockets due to the scarcity of labeled training data, while classical methods may produce ``blurry'' pockets~\cite{gervasoni2020comprehensive}. To demonstrate the ability of RAPID-Net to accurately detect both primary and secondary pockets, we focus on the therapeutically relevant case of the Nsp12 protein, an RNA-dependent RNA polymerase (RdRp), which plays a central role in viral RNA replication and is a key target for the development of antiviral drugs against SARS-CoV-2.

\noindent 
\textbf{RNA-dependent RNA Polymerase (RCSB PDB: 7BV2).} As illustrated in Fig.~\ref{Nsp12_Fig}, RAPID-Net successfully identifies the orthosteric remdesivir binding site~\cite{yin2020structural} -- containing residues 448, 545, 553, 555, 618, 759, 760, and 761 -- as its ``majority-voted'' pocket, consistent with predictions from other pocket prediction methods, PUResNet V1~\cite{kandel2021puresnet}, Kalasanty~\cite{stepniewska2020improving}, and PUResNet V2~\cite{jeevan2024puresnetv2}. In addition to the main binding site, computational studies have proposed allosteric~\cite{faisal2021computational} and secondary pockets~\cite{parvez2020prediction} that can be used for drug development. Unlike other pocket predictors, RAPID-Net also highlights putative allosteric and secondary pockets reported in the literature~\cite{faisal2021computational,parvez2020prediction}, demonstrating its ability to recover both primary and secondary binding sites.

Additionally, to illustrate RAPID-Net's ability to identify remote sites of therapeutic interest, we consider four proteins where such sites are well documented.

\noindent 
 \textbf{Thrombin (RCSB PDB: 1DWC).} The RCSB PDB entry 1DWC~\cite{banner1991crystallographic} is the crystal structure of human $\alpha$-thrombin in complex with the inhibitor MD-805. Thrombin plays a pivotal role in the coagulation cascade and is therefore a key target in treating acute coronary syndromes~\cite{weitz2002direct}. As shown in Fig.~\ref{1DWC_Fig}, Kalasanty~\cite{stepniewska2020improving} and PUResNet~\cite{kandel2021puresnet} predict binding pockets largely surrounding the ligand. By contrast, our model, RAPID-Net, detects an additional bulge extending towards residues 71, 73, 75, 76, and 77. These residues belong to the anion-binding Exosite~\rm{I}, which interacts with negatively charged substrates and cofactors such as fibrinogen, thrombomodulin, and COOH-terminal peptide of hirudin~\cite{crawley2007central,adams2006thrombin,bock2007exosites,troisi2021exosite,petrera2009long}. Our model's prediction, a ``tunnel'' connecting the active site to the Exosite~\rm{I} in Fig.~\ref{1DWC_Fig}, suggests possible long-range interactions, consistent with previous studies of long-range allosteric communication in thrombin~\cite{petrera2009long}.

\vspace{1em}
\noindent
\textbf{Human Carbonic Anhydrase I (RCSB PDB: 1AZM) and II (RCSB PDB: 3HS4).} Human carbonic anhydrase \rm{I} (hCA\rm{I}) in complex with a sulfonamide drug (PDB: 1AZM)~\cite{chakravarty1994drug} and hCA\rm{II} in complex with acetazolamide (AZM) (PDB: 3HS4)~\cite{sippel2009high} are key targets for glaucoma treatment~\cite{pfeiffer2001glaucoma} and diuretic therapy~\cite{supuran2008carbonic}, with broader potential for treating obesity, cancer, and Alzheimer’s disease~\cite{supuran2008carbonic}. In both isoforms, His64 acts as the primary proton shuttle~\cite{silverman1988catalytic,nair1991unexpected,lomelino2018crystallography}. 

As shown in Fig.~\ref{1AZM_Fig}, while PUResNet V1~\cite{kandel2021puresnet} and Kalasanty~\cite{stepniewska2020improving} predict the primary ligand binding region in 1AZM, our model additionally predicts a hook-shaped bulge wrapping around His64, suggesting an allosteric interaction.

In hCA\rm{II} (3HS4), a histidine cluster (His3, His4, His10, His15, His17, His64) mediates proton exchange between the active site and the environment~\cite{supuran2008carbonic}. Notably, 3HS4 contains two additional AZM molecules in shallow surface pockets--one previously identified~\cite{jude2006ultrahigh,srivastava2007structural} and another in a novel binding site~\cite{sippel2009high}.

As shown in Fig.~\ref{3HS4_1_Fig}, RAPID-Net identifies two majority-voted pockets: one around the catalytic site and another one near the histidine cluster. As shown in Fig.~\ref{3HS4_3_Fig}, the minority-reported pockets predicted by our model have $\text{PLI} = 92\%$ with these secondary binding poses in shallow pockets. This highlights the ability of our model to predict all binding sites in the protein, both primary and secondary.

In comparison, for this protein structure, Kalasanty~\cite{stepniewska2020improving} predicts only the main binding site, PUResNet V1~\cite{kandel2021puresnet} detects none, and PUResNet V2~\cite{jeevan2024puresnetv2} identifies only the interacting residues within the main binding site.

\vspace{1em}
\noindent
\textbf{Ls-AChBP (RCSB PDB: 2ZJV).} RCSB PDB ID 2ZJV~\cite{ihara2008lymnaea} represents Lymnaea stagnalis acetylcholine-binding protein (AChBP) bound to neonicotinoid clothianidin. AChBP serves as a model for nicotinic acetylcholine receptors (nAChRs) and their allosteric transitions~\cite{taly2009nicotinic}, providing insights into therapies for neurological disorders such as Alzheimer’s disease, schizophrenia, depression, attention deficit hyperactivity disorder, and tobacco addiction~\cite{arneric2007neuronal,levin2007nicotinic,romanelli2007central}. 

For the 2ZJV protein structure, neither Kalasanty~\cite{stepniewska2020improving} nor PUResNet V1~\cite{kandel2021puresnet} predict any pockets. However, the majority-voted pockets from our model and the interacting residues of PUResNet V2~\cite{jeevan2024puresnetv2} illustrated in Fig.~\ref{2ZJV_1_Fig} are predicted around all five true ligand binding positions.

Furthermore, our model's pockets include distinct bridges toward residues Thr155-Asp160 in the flexible F-loop, suggesting indirect ligand interactions. This finding aligns with the results from~\cite{ihara2008lymnaea}, who highlighted that induced-fit movements of loop regions, including the F-loop, are essential for ligand recognition by neonicotinoids.

These examples demonstrate our model's capability to identify regions influencing ligand binding--even without direct ligand contact. By revealing distal regions such as Exosite~\rm{I} in thrombin, His64 in hCA, and remote loop segments in Ls-AChBP, our approach provides key structural insights that can drive allosteric inhibitor design and broader therapeutic innovation.

\section{Discussion and Conclusions}
\label{Discussion}

As the number of protein structures without known binding sites continues to grow, performing binding-site-agnostic (or ``blind'') docking has become crucial for structure-guided drug design. Successful blind docking relies heavily on accurately identifying the search grid where the ligand is likely to bind. However, most existing pocket prediction tools operate irrespectively of docking pipelines and are evaluated using metrics that do not directly correlate with docking success. To address this gap, we developed RAPID-Net, an ML-based pocket identification tool specifically designed for seamless integration with docking pipelines. We tested RAPID-Net's effectiveness in guiding blind docking using AutoDock Vina 1.2.5~\cite{eberhardt2021autodock}, but our approach can be easily adapted to any docking software that requires a well-defined search grid.

When guided by RAPID-Net, AutoDock Vina~\cite{eberhardt2021autodock} outperforms DiffBindFR~\cite{zhu2024diffbindfr} by over 5\% in blind docking accuracy on the PoseBusters~\cite{Buttenschoen2024} dataset, highlighting the direct relationship between improved pocket identification and increased docking accuracy. Furthermore, RAPID-Net provides precise and compact search grids, enabling the docking of ligands to large proteins such as 8F4J from the PoseBusters~\cite{Buttenschoen2024} dataset, which tools like AlphaFold 3~\cite{abramson2024accurate} cannot process in whole, highlighting the critical importance of accurate and focused search areas for cost-effective docking.

We found that another major factor limiting docking accuracy is the reranking of generated poses. In addition to precise pocket identification, developing improved reranking tools could further enhance the overall performance of our combined scheme by more effectively selecting the most favorable poses from the generated ensemble.

We attribute the success of RAPID-Net to the following key changes and innovations we introduced: 

\begin{itemize} 

\item \textbf{Soft Labeling and ReLU Activation}: Unlike conventional binary segmentation tasks, we applied soft labels and ReLU activation in the output layer, drawing inspiration from medical image segmentation techniques. This approach enables the model to more effectively differentiate between the internal regions of pockets and their boundaries. 

\item \textbf{Attention Mechanism}: We integrated a single attention block within the encoder-decoder bottleneck, enhancing the model's ability to focus on relevant features while reducing the risk of overfit in inherently noisy data environments.

\item \textbf{Simplified Architecture}: By eliminating excessive residual connections, we streamlined the model architecture, resulting in improved performance. This simplification was effective given the noisy nature of the dataset.

\end{itemize}

Furthermore, RAPID-Net demonstrates the ability to identify  ``bridges'' to distal sites further than $6.5\ \text{\AA}$ from the primary pocket that are often critical for therapeutic intervention. This makes RAPID-Net a promising tool for targeting remote therapeutic sites and broadening the scope of structure-based drug design.

Additionally, beyond improved docking accuracy, in our future work, we plan to integrate RAPID-Net into de novo pocket-conditioned drug design pipelines by guiding the generation of novel compounds based on the shape features of the predicted pockets. Although our current study does not directly exploit these features, using the predicted region solely as a search box, as illustrated in Fig.~\ref{Vina_Setup_Fig}, they may prove highly valuable for generative models~\cite{cremer2024pilot,qiao20253d,alakhdar2024diffusion,zhang2023resgen,huang2024dual}.

\section*{Usage of Our Model and Reproducibility of Results}

All code and relevant data (proteins, pockets, poses) are publicly available on GitHub~\footnote{\href{https://github.com/BalytskyiJaroslaw/RAPID-Net}{https://github.com/BalytskyiJaroslaw/RAPID-Net}}. To reproduce our results, interested readers should download these files and follow the instructions provided in our repository.

To apply our trained model to predict pockets for other proteins and perform subsequent docking with AutoDock Vina~\cite{eberhardt2021autodock}, we provide an accompanying notebook with step-by-step guidance in the same repository. Additionally, our pocket predictor can be integrated with other docking tools, for example, by using spherical search grids. Users may also refine the ensemble of poses generated by AutoDock Vina~\cite{eberhardt2021autodock} guided by our pocket predictor by any preferred reweighting method, leveraging our pocket predictions to improve docking accuracy.

\section*{Author Contributions}

YB designed the neural network, conducted model training and docking experiments, and drafted the manuscript. IH and AB contributed to data analysis and provided medical expertise. CVK supervised the research and contributed to data analysis. All authors contributed to the proofreading and approved the final version of the manuscript. 

\section*{Acknowledgments}
Research reported in this publication was supported by the National Institute of Diabetes and Digestive and Kidney Diseases of the National Institutes of Health under award number R01DK076629. We are grateful to Richard J. Barber and the Barber Integrative Metabolic Research Program for financial support. We also thank James Granneman and Kelly McNear for valuable discussions.

\bibliography{apssamp.bib}

\end{document}